\documentstyle[rmp,harvard,preprint,aps]{revtex}
\citationstyle{dcu}
\citationmode{default}

\begin{document}
\title{Continuous Quantum Phase Transitions }
\author{S.\ L.\ Sondhi}
\address{Department of Physics, Princeton University, Princeton NJ
08544}

\author{S.\ M.\ Girvin}
\address{Department of Physics, Indiana University, Bloomington IN 47405}

\author{J.\ P.\ Carini}
\address{Department of Physics, Indiana University, Bloomington IN 47405}

\author{D. Shahar}
\address{Department of Electrical Engineering, Princeton University,
Princeton NJ 08544}

\maketitle

\vfill
\today
\newpage

\begin{abstract}
A quantum system can undergo a continuous phase transition at the absolute
zero of temperature as some parameter entering its Hamiltonian is varied.
These transitions are particularly interesting for, in contrast to their
classical finite temperature counterparts, their dynamic and static
critical behaviors are intimately intertwined. We show that considerable
insight is gained by considering the path integral description of the
quantum statistical mechanics of such systems, which takes the form of the
{\em classical} statistical mechanics of a system in which time appears as
an extra dimension. In particular, this allows the deduction of scaling
forms for the finite temperature behavior, which turns out to be described
by the theory of finite size scaling. It also leads naturally to the notion
of a temperature-dependent dephasing length that governs the crossover
between quantum and classical fluctuations. We illustrate these ideas 
using Josephson junction arrays and 
with a set of recent experiments on phase transitions in systems exhibiting
the quantum Hall effect.
\end{abstract}
%\pacs{}
\newpage

\tableofcontents
\newpage

% TEXT
%SLS

A century subsequent to Andrews's discovery of critical
opalescence\footnote{Opalescence is the strong reflection of
light by a system (such as an opal) due to fluctuations in its index of
refraction on length scales comparable to the wavelengths of visible light.
A liquid vapor system near its critical point has large density
fluctuations on length scales which can reach microns. This causes the
system, which is normally transparent, to have a cloudy appearance.}
in carbon dioxide, continuous phase transitions
continue to be a subject of great interest to physicists. The appeal
of the subject is twofold. First, the list of systems that exhibit interesting
phase transitions continues to expand; it now includes the Universe
itself! Second, the formal theory of equilibrium phase transitions has
found applications in problems as diverse as constructing field and
string theories of elementary particles, the transition to chaos in
dynamical systems, and the long time behavior of systems out of
equilibrium.

Our purpose in this Colloquium is to give a brief and qualitative account
of some basic features of a species of phase
transitions,\footnote{Henceforth we shall use
phase transitions as a shorthand for continuous phase transitions.}
termed `Quantum Phase Transitions' (QPTs), that have attracted much interest
in recent years.
These transitions take place at the absolute zero of temperature,
where crossing the phase boundary means that the {\em quantum ground state}
of the system changes in some fundamental way. This is accomplished by
changing not the temperature, but some parameter in the
Hamiltonian of the system. This parameter might be
the charging energy in Josephson junction arrays (which controls their
superconductor-insulator transition), the magnetic field in a quantum Hall
sample (which controls the transition between quantized Hall plateaus),
doping in the parent compound of a high T$_{\rm c}$ superconductor
(which destroys the antiferromagnetic spin order),
or disorder in a conductor near its metal-insulator transition
(which determines the conductivity at zero temperature). These and other QPTs
raise new and fascinating issues for theory and experiment,
most notably the inescapable necessity of taking quantum effects
into account.

Exactly what quantum effects are at issue is a bit subtle. As a
corollary of our definition, all finite temperature\footnote{In an
almost standard abuse of language, we refer to non-zero temperatures
as finite.} transitions are to be considered ``classical'', even in
highly quantum mechanical systems like
superfluid helium or superconductors. It is not that quantum mechanics
is unimportant in these cases, for in its absence there would not {\em be}
an ordered state, i.e. the superfluid or superconductor. Nevertheless,
sufficiently close to the critical point quantum {\em fluctuations}
are important at the microscopic scale, but not at the longer length scales
that control the critical behavior; in the jargon of statistical
mechanics, quantum mechanics is needed for the existence of
an order parameter\footnote{An order parameter is a quantity which is
zero in the disordered phase and
non-zero in the ordered state. In systems that spontaneously break some
symmetry in the ordered state, the nature (and value) of the order parameter
reflects this broken symmetry. Thus for example in an Ising ferromagnet, the
magnetization is a positive or negative real number indicating the
difference in populations of the up and down spins. See Goldenfeld (1992).}
 but it is classical thermal fluctuations that govern it
at long wavelengths.
For instance, near the superfluid `lambda' transition in $^4$He,
the order parameter is a complex-valued field which is related to the
underlying condensate wave function. However, its critical fluctuations
can be captured exactly by doing {\em classical} statistical mechanics
with an effective Hamiltonian for the order parameter
field (for instance the phenomenological Landau-Ginsburg free energy
functional \cite{goldenfeld}).
%%As a consequence of universality for classical critical phenomena, this
%%effective classical system can be replaced by a much simpler one,
%%here the 3D XY model, which has precisely the same critical behavior
%%(exponents, scaling functions etc) as the real quantum
%%system. (The important fluctuations of the complex
%%$^4$He order parameter involve,
%%not its magnitude but rather its variable phase. The angular variables of
%%the XY model capture this feature.)

The physics behind the classical nature of finite temperature transitions
is the following: Phase transitions are quite generally accompanied
by a divergent correlation length and correlation time, i.e. the
order parameter (e.g. the magnetization in a ferromagnet) fluctuates
coherently over increasing distances and ever more slowly.
%(That is, the fluctuations are correlated over larger distances and times.)
The latter implies that there is a frequency $\omega^*$ associated
with the critical fluctuations that vanishes at the transition.
A quantum system behaves classically if the temperature exceeds
all frequencies of interest, and since $\hbar \omega^* \ll k_B T_c$
close to the transition, the critical fluctuations will behave classically.

This argument also shows that the case of QPTs where $T_c=0$, is qualitatively
different and that there the critical fluctuations must be treated
quantum mechanically. In the following we will describe
the language and physical pictures that enable such a treatment and
which have come into common usage among practitioners in the field
in the last few years. 
Although much of this wisdom, which has its roots in work on quantum 
Ising models \cite{young,suzuki}, dates back to the work of 
Hertz (1976)\footnote{We should note that the contemporaneous explosion
of work on the one-dimensional electron gas \cite{emery} provided
important, early illustrations of these ideas.}, it remains unknown or 
poorly understood in the wider community and extracting it from the 
literature remains a daunting task. It is our hope here to communicate 
this set of ideas to a wider audience with the particular desire to be 
helpful to newcomers to this field, experimentalists and theorists alike. 

Our discussion is organized as follows. In Section~\ref{qsm} we introduce
the statistical mechanics of quantum systems and the path
integral \cite{feynman} approach to it, which is an extremely useful
source of intuition in these problems. A running theme throughout this
discussion is the intertwining of dynamics and thermodynamics in quantum
statistical mechanics. In Section~\ref{pfpi} we describe the general
features of a QPT at $T=0$ and how a non-zero temperature alters the
physics. This leads naturally to a discussion of what kind of
scaling behavior in experiments is evidence of
an underlying QPT in Section~\ref{e1djja}. We illustrate this using
particular examples from phase transitions in quantum Hall systems. We
end in Section IV with a brief summary and pointers to work on QPTs in
other interesting systems. Readers interested in a highly informative
discussion at a higher technical level should consult the recent
beautiful review article by Sachdev (1996).

\section{Quantum Statistical Mechanics: Generalities}
\label{qsm}

%%\noindent
%%{\em essential buzz-ideas: thermodynamics is imaginary time
%%evolution, inverse temperature is system size in time, dynamics is
%%thermodynamics, coupling is temperature}

Before we discuss what happens in the vicinity of a QPT, let us recall some
very general
features of the statistical mechanics of quantum systems. The quantities
of interest are the partition function of the system,
\begin{equation}
Z(\beta) = {\rm Tr}\,\, e^{-\beta H}
\end{equation}
and the expectation values of various operators,
\begin{equation}
\langle O \rangle = \frac{1}{Z(\beta)}{\rm Tr}\,\, O e^{-\beta H}.
\end{equation}
In writing these formal expressions we have assumed a finite temperature,
$k_B T = 1/\beta$. To get at what happens exactly at $T=0$ we take the
$T \rightarrow 0$ limit. Upon doing so, the free energy,
$F = - \frac{1}{\beta}\ln{Z}(\beta)$, becomes the ground state energy and
the various thermal averages become ground state expectation values.
\null From $Z$ we can get all the thermodynamic quantities of interest.
Expectation values of operators
of the form $O \equiv A({\bf r} t) A({\bf r'} t')$ are related to the
results of dynamical scattering and linear response measurements. For
example, $A$ might be the local density (X-ray scattering) or current
(electrical transport).

\subsection{Partition Functions and Path Integrals}
\label{pfpi}

Let us focus for now on the expression for $Z$. Notice that the operator
density matrix, $e^{-\beta H}$, is the same as the time evolution operator,
$e^{-iH{\cal T}/\hbar}$,
provided we assign the {\em imaginary} value ${\cal T} = - i \hbar\beta$ to the
time interval over which the system evolves. More precisely, when
the trace is written in terms of a complete set of states,
\begin{equation}
Z(\beta) = \sum_n \langle n | e^{-\beta H} | n \rangle,
\label{eq:Zimag}
\end{equation}
$Z$ takes the form
of a sum of imaginary time transition amplitudes for the system to
start in some state $|n\rangle$ and
{\em return to the same state} after an imaginary time interval $-i\hbar\beta$.
Thus we see that calculating the thermodynamics
of a quantum system is the same as calculating transition amplitudes
for its evolution in imaginary time, the total time interval being fixed
by the temperature of interest. The fact that the time interval happens to
be imaginary is not central. The key idea we hope to transmit to the
reader is that Eq.(\ref{eq:Zimag}) should evoke an image of quantum
dynamics and temporal propagation.

This way of looking at things can be given a particularly beautiful
and practical
implementation in the language of Feynman's path integral formulation
of quantum mechanics \cite{feynman}.
 Feynman's prescription is that the net transition
amplitude between two states of the system can be calculated by summing
amplitudes for all possible paths between them.
%%%%%%%%%%%%%%%%%%%%%%%%%%%%%%%%%%%%%%%%%%%%%%%%%%%%%%%%%%%%%%%%%%%%%%%%
%%More precisely, if $x=a$
%%and $x=b$ are the states at time $t=0$ and $t=\tau$
%%($x$ represents some set of
%%co-ordinates for the system) then we sum the amplitudes $e^{i S[x(t)]/\hbar}$
%%where $S[x(t)]$ is the classical action for the path $x(t): x(0)=a {\rm
%%and} x(\tau)=b$, and equals the integral of the Lagrangian
%%$\int {\it dt} L(x,\dot x)$ over the path. It follows then that
%%$Z(\beta)$ can be written as a path integral in imaginary time $t=-i \tau$:
%%\begin{equation}
%%Z(\beta) = \int_{x(0)=x(\hbar\beta)} {\cal D}x(\tau)\;
%%e^{-\frac{1}{\hbar} \int_{0}^{\hbar\beta} d\tau\;
%%L_E(x,\dot x)},
%%\label{eq15}
%%\end{equation}
%%where $L_E$ is the Euclidean Lagrangian $L_E = - L(t \rightarrow -i \tau)$.
%%
%%%%%%%%%%%%%%%%%%%%%%%%%%%%%%%%%%%%%%%%%%%%%%%%%%%%%%%%%%%%%%%%%%%%%%%%%%
The path taken by the system is defined by specifying the state of the
system at a sequence of finely spaced intermediate time steps. Formally we
write
\begin{equation}
 e^{-\beta H} =
\left[e^{-\frac{1}{\hbar}\delta\tau H} \right]^N,
\end{equation}
where $\delta\tau$ is a time interval\footnote{For convenience we have
chosen $\delta\tau$ to be real, so that the small interval of imaginary
time that it represents is $-i\delta\tau$.}
 which is small on the time scales
of interest ($\delta\tau = \hbar/\Gamma$ where $\Gamma$ is some ultraviolet
cutoff) and
$N$ is a large integer chosen so that $N\delta\tau = \hbar\beta$.
We then insert a sequence of sums over complete sets of
intermediate states into
the expression for $Z(\beta)$:
\begin{equation}
Z(\beta) = \sum_n \sum_{m_1,m_2,\ldots,m_N}
\langle n |
e^{-\frac{1}{\hbar}\delta\tau H}
|m_1\rangle\langle m_1|
e^{-\frac{1}{\hbar}\delta\tau H}
|m_2\rangle\langle m_2|
\dots
|m_N\rangle\langle m_N|
e^{-\frac{1}{\hbar}\delta\tau H}
| n \rangle.
\label{eq:complete_set}
\end{equation}
This rather messy expression actually has a rather simple physical
interpretation. Formally inclined readers will
%%[formally horizontal readers may skip this paragraph]
observe that the expression for the {\em quantum} partition function in
Eq.~(\ref{eq:complete_set}) has the form of a {\em classical} partition
function, i.e. a sum over configurations expressed in terms of a transfer
matrix, if we think of imaginary time as an additional
spatial dimension. In particular, if our quantum system lives in $d$
dimensions,
the expression for its partition function looks like a classical partition
function for a system with $d+1$ dimensions, except that the extra dimension
is finite in extent---$\hbar \beta$ in units of time. As $T \rightarrow 0$
the system size in this extra ``time'' direction diverges and we get
a truly $d+1$ dimensional, effective, classical system.

Since this equivalence
between a $d$ dimensional quantum system and a $d+1$ dimensional classical
system is crucial to everything else we have to say,
and since Eq.~(\ref{eq:complete_set}) is probably not very illuminating for
readers not used to a daily regimen of transfer matrices,
 it will be very
useful to consider a specific example in order to be able to visualize what
Eq.~(\ref{eq:complete_set}) means.

\subsection{Example: 1D Josephson Junction Arrays}
\label{e1djja}

Consider a one-dimensional array comprising a large number $L$
of identical Josephson junctions as illustrated in
Fig.~(\ref{fig:JJdiagram}).  Such arrays have recently been studied by
by Haviland and Delsing. \cite{haviland}
 A Josephson junction is a tunnel junction
connecting two superconducting metallic grains. Cooper pairs of electrons
are able to tunnel back and forth between the grains and hence communicate
information about the quantum state on each grain. If the Cooper pairs are
able to move freely from grain to grain throughout the array, the system is
a superconductor. If the grains are very small however, it costs a large
charging energy to move an excess Cooper pair onto a grain. If this energy
is large enough, the Cooper pairs fail to propagate and become stuck on
individual grains, causing the system to be an insulator.

The essential degrees of freedom in this
system are the phases of the complex superconducting order parameter on the
metallic elements connected by the junctions\footnote{It is believed that
neglecting fluctuations in the magnitude of the order parameter is a good
approximation; see Bradley and Doniach (1984); Wallin {\it et al.} (1994)} 
and their conjugate variables, the charges (excess Cooper pairs, or 
equivalently the voltages) on
each grain. The intermediate state $|m_j\rangle$ at time $\tau_j \equiv
j\delta\tau$, that enters the quantum partition function
Eq.~(\ref{eq:complete_set}), can thus be defined by specifying
the set of phase angles $\{\theta(\tau_j)\} \equiv
[\theta_1(\tau_j),\theta_2(\tau_j),\ldots,\theta_L(\tau_j)]$.
Two typical paths or time histories
on the interval $[0,\hbar\beta]$ are illustrated in
Fig.~(\ref{fig:JJpath}) and Fig.~(\ref{fig:JJpath2}),
where the orientation of the arrows (`spins') indicates the local
phase angle of the order parameter.
The statistical weight of a given path, in the sum in
Eq.~(\ref{eq:complete_set}), is given by the
product of the matrix elements
\begin{equation}
\prod_j\langle \{\theta(\tau_{j+1})\} |
e^{-\frac{1}{\hbar}\delta\tau H} |\{\theta(\tau_j)\}\rangle,
\label{eq:statwgt}
\end{equation}
where
\begin{equation}
H = \frac{C}{2} \sum_j V_j^2 - E_{\rm J} \cos\left(\hat\theta_j -
\hat\theta_{j+1}\right),
\label{eq:H6}
\end{equation}
is the quantum Hamiltonian of the Josephson junction
array.
Here $\hat\theta_j$ is the operator representing
the phase of the superconducting order parameter on the $j$th
grain\footnote{Our notation here is that
$\{\theta(\tau)\}$ refers to the configuration of the entire set of angle
variables at time slice $\tau$. The $\hat\theta$'s appearing in the
Hamiltonian in Eq.(\ref{eq:H6}) are angular coordinate operators and $j$ is
a site label. The state at time slice $\tau$ is an eigenfunction of these
operators:
$\cos\left(\hat\theta_j -
\hat\theta_{j+1}\right)|\{\theta(\tau)\}\rangle
= \cos\left(\theta_j(\tau) -
\theta_{j+1}(\tau)\right)|\{\theta(\tau)\}\rangle.$};
$V_j \equiv -i\frac{2e}{C}\frac{\partial}{\partial\theta_j}$
is conjugate to the phase\footnote{It is useful to think of this as a
quantum rotor model. The state with wave function $e^{im_j\theta_j}$
has $m_j$ units of angular momentum representing $m_j$ {\em excess}
Cooper pairs on grain $j$.
The Cooper-pair number operator in this representation is
$n_j = - i\frac{\partial}{\partial\theta_j}$. See \cite{wallinetalPRB}.
The cosine term in Eq.(\ref{eq:H6}) is a `torque' term which transfers units of
conserved angular momentum (Cooper pairs) from site to site. Note that
the potential energy of the bosons is represented,
somewhat paradoxically, by the kinetic energy of the quantum rotors 
and vice versa.}
 and is the voltage on the $j$th junction, and
$E_{\rm J}$ is the Josephson coupling energy.
The two terms in the Hamiltonian
then represent the charging energy of each grain
and the Josephson coupling of the phase across the junction between grains.

As indicated previously, we can map the quantum statistical mechanics of the 
array onto  classical statistical mechanics by identifying the the statistical 
weight of a space-time path in Eq.~(\ref{eq:statwgt}) with the Boltzmann weight 
of a two-dimensional spatial configuration of a classical system. 
In this case the classical system is therefore a {\em two-dimensional} 
X-Y model, i.e. its degrees of freedom are planar spins, specified by angles 
$\theta_i$, that live on a two-dimensional square lattice. (Recall that at temperatures
above zero, the lattice has a finite width $\hbar \beta/\delta \tau$ in the 
temporal direction.) While the degrees
of freedom are easily identified, finding the classical hamiltonian for this
X-Y model is somewhat more work and requires an explicit evaluation of the
matrix elements which interested readers can find in the Appendix.

It is shown in the Appendix
that, in an approximation that preserves the universality
class of the problem\footnote{That is, the approximation is such that the 
universal aspects of the critical behavior such as the exponents and scaling
functions will be given without error. However, non-universal quantities such
as the critical coupling will differ from an exact evaluation. Technically,
the neglected terms are irrelevant at the fixed point 
underlying the transition.},
the product of matrix elements in Eq.~(\ref{eq:statwgt}) can be rewritten in
the form $e^{-H_{\rm XY}}$ where the Hamiltonian of the equivalent classical 
X-Y model is 
\begin{equation}
H_{\rm XY} =
\frac{1}{K}\sum_{\langle ij\rangle} \cos(\theta_i - \theta_j),
\label{eq:equivXY}
\end{equation}
and the sum runs over near-neighbor points in the {\em two-dimensional}
 (space-time)
lattice.\footnote{Notice this crucial change in notation from Eq.(\ref{eq:H6})
where $j$ refers to a point in 1D space, not 1+1D space-time.} The nearest
neighbor character of the couplings identifies the classical model as {\em the}
2D X-Y model, extensively studied in the context of 
superfluid and superconducting
films \cite{goldenfeld,chaikin-lubensky}. 
We emphasize that while the straightforward
identification of the degrees of freedom of the classical model 
in this example is robust, this simplicity of the resulting classical 
Hamiltonian is something of a minor miracle.

It is essential to note that the dimensionless
coupling constant $K$ in $H_{\rm XY}$, which plays the role of the 
temperature in the classical model, depends on the ratio of the
capacitive charging energy $E_{\mathrm C} = \frac{(2e)^2}{C}$
to the Josephson coupling $E_{\rm J}$ in the array,
\begin{equation}
K \sim \sqrt{\frac{E_{\mathrm C}}{E_{\rm J}}}.
\end{equation}
and has nothing to do with the physical temperature. (See  Appendix.)
The physics here is that a large Josephson coupling produces a small value
of $K$ which
favors coherent ordering of the phases. That is, small $K$ makes it
unlikely that $\theta_i$ and $\theta_j$ will differ significantly, even
when sites $i$ and $j$ are far apart (in space and/or time).
Conversely,
a large charging energy leads to a large value of $K$ which favors
zero-point fluctuations of the phases and disorders the system.
That is, large $K$
means that the $\theta$'s are nearly independent and all values are
nearly equally likely.\footnote{Because particle number is conjugate to the
phase [$\hat n_j = -i \frac{\partial}{\partial\theta_j}$], a state of
indefinite phase on a site has definite charge on that site,
as would be expected for an insulator.}
Finally, we note that this
equivalence generalizes to d-dimensional arrays and d+1-dimensional
classical XY models.

\subsection{Quantum-Classical Analogies}

This specific example of the equivalence between a quantum
system and a classical system with an extra `temporal' dimension,
illustrates several general correspondences between quantum systems
and their effective classical analogs.

Standard lore tells us that the classical XY model has an order-disorder
phase transition
as its temperature $K$ is varied. It follows that the quantum array
has a phase transition as the ratio of its charging and Josephson
energies is varied.
One can thus see why it is said that
the superconductor-insulator quantum phase transition in a 1-dimensional
Josephson junction array is in the same universality class as the
order-disorder phase transition of the 1+1-dimensional classical XY
model.  [One crucial caveat is that the XY model universality class has
strict particle-hole symmetry for the bosons (Cooper pairs) on each 
site.  In reality, Josephson junction arrays contain
random `offset charges' which destroy this symmetry and change the
universality class \cite{wallinetalPRB}, a fact which is all too often 
overlooked.]

We emphasize again that $K$ is the temperature only in the effective
classical problem. In the quantum case,
the physical temperature is presumed to be nearly zero
and only enters as the finite size of the system in the imaginary time
direction. {\em The coupling constant $K$, the fake `temperature,' is a
measure not of thermal fluctuations, but of the strength of quantum
fluctuations, or zero point motion of the phase variables}.\footnote{Zero
point motion of the phase variables is caused by the fact that there is an
uncertainty relation between the phase and the number of Cooper pairs on a
superconducting grain. The more well-defined the phase is, the larger the
uncertainty in the charge is. This costs capacitive energy.} This
notion is quite confusing, so the reader might be well advised to pause
here and contemplate it further. It may be useful to examine
Fig.~(\ref{fig:two-sizes}), where we show a space time lattice for the XY
model corresponding to a Josephson junction array at a certain
temperature, and at a temperature half as large. The size of the lattice
constant in the time direction [$\delta\tau$ in the path integral in
Eq.~(\ref{eq:complete_set})] and $K$ are the {\em same} in both cases
even though the physical temperature is not the same. The
only difference is that one lattice is larger in the time direction than
the other.

In developing intuition about this picture, it may be helpful to see how
classical physics is recovered at very high temperatures. In that limit,
the time interval $\hbar\beta$ is very short compared to the periods
associated with the natural
frequency scales in the system and typical time histories will consist of
a single static configuration which is the same at each time slice.
The dynamics therefore drops out of the problem and a Boltzmann weight
$\exp(-\beta H_{\rm classical})$ is recovered from the path integral.

The thermodynamic phases of the array can be identified from those of the
XY model. A small value of
$K$ corresponds to low temperature in the classical system and so
the quantum system will be in the ordered ferromagnetic phase of the XY
model, as illustrated in Fig.~(\ref{fig:JJpath}). There will be long-range
correlations in both space and time of the phase variables.\footnote{In
this special 1+1D case, the correlations happen to decay algebraically
rather than being truly of infinite range.} This indicates that the
Josephson coupling dominates over the charging energy, and the order
parameter is not fluctuating wildly in space or time so that the system is
in the {\em superconducting} phase. For large $K$, the system is disordered
and the order parameter fluctuates wildly. The correlations decay
exponentially in space and time as illustrated in
Fig.~(\ref{fig:JJpath2}). This indicates that the system is in the {\em
insulating} phase, where the charging energy dominates over the Josephson
coupling energy.

Why can we assert that correlations which decay
exponentially in imaginary time indicate an excitation gap characteristic
of an insulator? This is readily seen by noting that the Heisenberg
representation of an operator in imaginary time is
\begin{equation}
A(\tau) = e^{H\tau/\hbar} A e^{-H\tau/\hbar}
\end{equation}
and so the (ground state) correlation function for any
operator can be expressed in terms of a complete set of states as
\begin{equation}
G(\tau) \equiv \langle 0|A(\tau)A(0)|0\rangle =
\sum_m e^{-(\epsilon_m - \epsilon_0)\tau/\hbar} |\langle 0|A|m\rangle|^2,
\label{eq:qmcorr}
\end{equation}
where $\epsilon_m$ is the energy of the $m$th excited state.
The existence of a finite minimum excitation gap
$\Delta_{01} \equiv \epsilon_1 - \epsilon_0$
guarantees that for long (imaginary)
times the correlation function will decay
exponentially,\footnote{At $T\ne 0$ and for very long times (comparable to
$\hbar\beta$), the finiteness in the time direction will modify this result.
Also, we implicitly assume here that $\langle 0|A|0\rangle=0$.} i.e.,
\begin{equation}
G(\tau) \sim e^{-\Delta_{01}\tau/\hbar} \ .
\end{equation}

%\noindent
%STEVE: LENGTH DEPENDENCE OF RESISTANCE IN SC PHASE?

To recapitulate, we have managed to map the finite temperature 1D quantum
problem into
a 2D classical problem with one finite dimension that diverges as
$T \rightarrow 0$. The parameter that controls the fluctuations in the
effective classical problem does {\em not} involve $T$, but instead
is a measure of the quantum fluctuations. The classical model exhibits
two phases, one ordered and one disordered. These correspond to the
superconducting and insulating phases in the quantum problem. In the
former the zero-point or quantum fluctuations of the order parameter are
small. In the latter they are large.
The set of analogies developed here between quantum and classical critical
systems is summarized in Table~\ref{tableA}.

Besides the beautiful formal properties of the analogy between
the quantum path integral and $d+1$ dimensional statistical mechanics,
there are very practical advantages to this analogy. In many cases,
particularly for systems without disorder, the universality class of
the quantum transition is one that has already been
studied extensively classically and a great deal may already be known about
it. For new universality classes, it is possible to do the quantum
mechanics by classical Monte Carlo or molecular dynamics simulations of the
appropriate $d+1$-dimensional model.

Finally, there is a special feature of
our particular example that should be noted. In this case the quantum
system, the 1D Josephson junction array (which is also the 1D quantum X-Y
model), has mapped onto a classical model in which space and time enter
in the same fashion, i.e., the isotropic
2D classical X-Y model. Consequently, the dynamical exponent $z$ (to be
defined below) is unity. This is not true in general---depending upon
the quantum kinetics,
the coupling in the time direction can have a very different
form and the effective classical system is then intrinsically anisotropic
and not so simply related to the starting quantum system.

\subsection{Dynamics and Thermodynamics}

We end this account of quantum statistical mechanics by commenting
on the relationship between dynamics and thermodynamics.
In classical statistical mechanics, dynamics and thermodynamics are
separable, i.e., the momentum and position sums in the partition
function are totally independent. For example, we do not need to know
the mass of the particles to compute their positional correlations.
In writing down simple non-dynamical models, e.g. the
Ising model, we typically take advantage of this simplicity.

This freedom is lost in the quantum problem because coordinates and momenta
do not commute.\footnote{Stated more formally,
calculating $Z$ classically only requires knowledge
of the form of the Hamiltonian function and not of the equations of
motion, while both enter the quantum calculation. Recall that
$H$ alone does not fix the equations of motion; one also needs
the Poisson brackets/commutators among the phase space variables. While
these determine the classical oscillation frequencies, they
do not enter the classical calculation of $Z$. In quantum mechanics
$\hbar$ times the classical oscillation frequencies yields the energy level
spacing. Hence the commutators are needed to find the eigenvalues of
the quantum $H$ needed to compute the quantum $Z$.}
It is for this reason that our path integral expression for $Z$
contains information on the imaginary time evolution of the system
over the interval $[0,\hbar\beta]$,
and, with a little bit of care, that information can be used to get the dynamics
in real time by the analytic continuation,
\begin{equation}
G(\tau) \longrightarrow G(+it)
\end{equation}
in Eq.~(\ref{eq:qmcorr}).
Stating it in reverse, one cannot solve for the
thermodynamics without also solving for the dynamics---a feature that
makes quantum statistical mechanics more interesting but that much
harder to do!

Heuristically, the existence of $\hbar$ implies that energy scales that
enter thermodynamics necessarily determine time scales which then enter
the dynamics and vice-versa. Consider the effect of a characteristic
energy scale, such as a gap $\Delta$, in the spectrum. By the uncertainty
principle
there will be virtual excitations across this gap on a time scale
$\hbar/\Delta$, which will appear as the characteristic time scale for
the dynamics. Close to the critical point, where $\Delta$ vanishes, and
at finite temperature this argument gets modified---the relevant uncertainty
in the energy is now $k_BT$ and the characteristic time scale is
$\hbar\beta$. In either case, the linkage between dynamics and
thermodynamics is clear.

%SLS96 I think this is not needed for most readers.
%This raises another common confusion
%between imaginary time and computer time.\footnote{Not to be confused
%with computer `real' time.}
%It is well known that classical systems (and their simulations) suffer
%`critical slowing down' near critical points. In a molecular dynamics
%simulation of a classical system (using realistic dynamics), computer time
%and real time are more or less the same thing. We can compute dynamical
%correlation functions for some quantity by measuring its value at some
%(computer) time $t'$ and a later time $t$ and averaging
%$\langle {\cal O}(t){\cal O}(t')\rangle$. This is {\em not} what we do to
%get dynamical correlations for the quantum system. We perform the
%simulation and measure the static or equal (computer) time correlation
%function $\langle {\cal O}({\bf r}) {\cal O}({\bf r}')\rangle$ where
%${\bf r}$ and ${\bf r}'$ are different points in the space-time lattice.
%As noted above, quantum dynamics is measured by the dependence of
%the correlation function on the separation of ${\bf r}$ and ${\bf r}'$
%in the `time' direction.

%As a practical consideration however, we also have to deal with the
%critical slowing down in computer time. Since we are interested in the
%quantum critical point, the d+1-dimensional classical system we are
%simulating is automatically at or near its critical point and so the
%simulation automatically suffers from critical slowing down in computer
%time.

\section{Quantum Phase Transitions} \label{qpt}

We now turn our attention to the immediate neighborhood of a
quantum critical point. In this region the mapping of the quantum
system to a d+1 dimensional classical model will allow us
to make powerful general statements about the former using
the extensive lore on critical behavior in the latter. Hence
most of the following will consist of a reinterpretation of
standard ideas in classical statistical mechanics in terms appropriate
for $d+1$ dimensions, where the extra dimension is imaginary time.

\subsection{$T=0$: Dynamic Scaling}

In the vicinity of a continuous
quantum phase transition we will find several features of
interest. First, we will find a correlation length that diverges as
the transition is approached.
That diverging correlation lengths are a generic feature
of classical critical points, immediately tells us that diverging lengths
and diverging {\em times} are automatically a
generic feature of quantum critical points, since one of the directions in
the d+1 dimensional space is time. This makes sense from the
point of view of causality. It {\em should} take a longer and longer time to
propagate information across the distance of the correlation length.

Actually, we have to be careful---as
we remarked earlier, the time direction might easily involve a different
set of interactions than the spatial directions, leading to a
distinct correlation ``length'' in the time direction. We will call the
latter $\xi_\tau$, reserving the symbol $\xi$ for the spatial correlation
length.
Generically, at $T=0$ both $\xi(K)$ and $\xi_\tau(K)$
diverge as $\delta \equiv K - K_c \longrightarrow 0$ in the
manner,\footnote{Here and in the following, we do not write the
microscopic length and time scales that are needed to make dimensional
sense of these equations. See Goldenfeld (1992).}
\begin{eqnarray}
\xi &\sim& |\delta|^{-\nu} \nonumber \\
\xi_\tau &\sim& \xi^z.
\end{eqnarray}
These asymptotic forms serve to define the correlation length exponent $\nu$,
and the {\em dynamical scaling} exponent, $z$.
The nomenclature is historical, referring to the extension of scaling
ideas from the study of static classical critical phenomena to dynamics
in the critical region associated with critical slowing down
\cite{hh1,hh2}. In the classical problem the extension was
a non-trivial step, deserving of a proper label. As remarked
before, the quantum problem involves statics and dynamics on the
same footing and so nothing less is possible. For the case of the
Josephson junction array considered previously, we found the simplest
possible result, $z=1$. As noted however this is a special isotropic
case and in general, $z \ne 1$.

As a consequence of the diverging $\xi$ and $\xi_\tau$, it turns out
that various physical quantities in the critical region close to the
transition have ({\em dynamic}) {\em scaling forms}, i.e. their dependence
on the independent variables involves homogeneity relations of the form:
\begin{equation}
{\cal O}(k, \omega, K) = \xi^{d_{\cal O}} O(k \xi, \omega \xi_\tau)
\label{eq:homogen}
\end{equation}
where $d_{\cal O}$ is called the scaling dimension\footnote{The
scaling dimension describes how physical quantities change under a
renormalization group transformation in which short wavelength degrees
of freedom are integrated out. As this is partly a naive change of scale,
the scaling dimension is often close to the naive (``engineering'')
dimension of the observable but (except at special, non-generic,
fixed points) most
operators develop ``anomalous'' dimensions. See Goldenfeld (1992).}
of the observable ${\cal O}$ measured at wavevector $k$ and frequency $\omega$.
The meaning of (and assumption behind) these scaling forms is
simply that, close to the critical point, there is no characteristic
length scale other than $\xi$ itself\footnote{For a more precise statement
that includes the role of cutoff scales, see Goldenfeld (1992).}
and no characteristic time scale other than $\xi_\tau$. Thus the specific
value of the coupling $K$ does not appear explicitly on the RHS of
Eq.(\ref{eq:homogen}). It is present only implicitly through the $K$
dependence of $\xi$ and $\xi_\tau$.

If we specialize to the scale invariant critical point, the scaling form
in Eq.(\ref{eq:homogen}) is no longer
applicable since the correlation length and times have diverged to infinity.
In this case the only characteristic length left is the wavelength 
$2\pi/k$ at which
the measurement is being made, whence the only characteristic frequency is
$\bar\omega \sim k^z$. As a result we find the simpler scaling form:
\begin{equation}
{\cal O}(k, \omega, K_c) = k^{-d_{\cal O}} \tilde{O}(k^z/\omega),
\label{eq:homogen_crit}
\end{equation}
reflecting the presence of {\em quantum} fluctuations on all length and time
scales.\footnote{Equivalently, we could have argued that the scaling
function on the RHS of Eq.(\ref{eq:homogen}) must for large arguments
$x,y$ have the form
$O(x,y) \sim x^{-d_{\cal O}} \tilde{\cal O}(x^z y^{-1})$ in order for the 
observable to have a sensible limit as the critical point is approached.}

The utility and power of these scaling forms can be illustrated by
the following example. In an ordinary classical system at a critical point
in $d$ dimensions
where the correlation length has diverged, the correlations of many
operators typically fall off as a power law
\begin{equation}
\tilde G(r) \equiv
\langle{\cal O}({\bf r}){\cal O}({\bf 0})\rangle \sim
\frac{1}{r^{(d-2+\eta_d)}},
\end{equation}
so that the Fourier transform diverges at small wavevectors like
\begin{equation}
G(k) \sim k^{-2+\eta_d}.
\end{equation}
Suppose that we are interested in a QPT for which the d+1-dimensional
classical system is effectively isotropic and the dynamical exponent $z=1$.
Then the Fourier transform of the correlation function for the
$d+1$-dimensional problem is
\begin{equation}
G(k,\omega_n) \sim \left[\sqrt{k^2 + \omega_n^2}\right]^{-2+\eta_{d+1}},
\end{equation}
where the $d+1$ component of the `wavevector' is simply the Matsubara
frequency used to Fourier transform in the time direction. Analytic
continuation to real frequencies via the usual prescription \cite{Mahan}
 $i\omega_n \longrightarrow \omega + i\delta$ yields the retarded
correlation function
\begin{equation}
G_{\rm R}(k,\omega + i\delta) \sim
\
\left[{k^2 - (\omega + i\delta)^2}\right]^{(-2+\eta_{d+1})/2}.
\label{eq:branchcut}
\end{equation}
Instead of a pole at the frequency of some coherently oscillating
collective mode, we see instead
that $G_{\rm R}(k,\omega + i\delta)$ has a branch cut for frequencies above
$\omega = k$ (we have implicitly set the characteristic velocity to unity).
Thus we see that there is no characteristic frequency other than $k$
itself (in general, $k^z$ as in Eq.(\ref{eq:homogen_crit})),
as we discussed above.
This implies that collective modes have become
overdamped and the system is in an incoherent diffusive regime. The review
by Sachdev contains some specific examples which nicely illustrate these
points \cite{sachdevIUPAP}.

Finally, three comments are in order. First, as we saw in the example
of the Josephson junction array,
a finite temporal correlation length
means that there is a gap in the spectrum of the quantum problem.
Conversely, critical systems are gapless.
Second, the exponent $z$ is a measure of
how skewed time is, relative to space, in the {\em critical} region.
This does not, {\em a priori}, have anything to do with what happens in
either of the phases. For example, one should resist the temptation to
deduce the value of $z$ via $\omega \sim q^z$
from the dispersion of any Goldstone mode\footnote{A Goldstone mode
is a gapless excitation that is present as a result of a broken continuous
symmetry in the ordered phase of a system. Broken continuous symmetry
means that the energy is degenerate under a continuous family of uniform
{\em global} symmetry transformations, for example uniform rotation of the
magnetization in an XY magnet. This implies that non-uniform but long
wavelength rotations must cost very little energy and hence there exists a
low-energy collective mode in which the order parameter fluctuates at long
wavelengths. See Goldenfeld (1992) and Chaikin and Lubensky (1995).}
 in the ordered phase. This is incorrect since the exponent $z$ is a
property of the critical point itself, not of the ordered phase.
Third, we should restate the well known wisdom that
the diverging lengths and the associated scaling of physical quantities
are particularly interesting because they represent {\em universal}
behavior, i.e., behavior
insensitive to microscopic details within certain global constraints such
as symmetry and dimensionality \cite{goldenfeld}.

\subsection{$T \ne 0$: Finite Size Scaling}

So far we have described the framework, appropriate to the system at
$T=0$, that would describe the underlying QPT in any system.
As the experimentally accessible behavior of the system necessarily
involves a non-zero temperature, we need to understand how to modify the
scaling forms of the previous section for the $T \ne 0$ problem.

The crucial observation for this purpose is , as noted earlier and
illustrated in Fig.~(\ref{fig:two-sizes}), that the {\em only} effect of
taking $T \ne 0$ in the partition function (\ref{eq:complete_set})
is to make the
temporal dimension {\em finite}; in particular, there is no change in the
coupling $K$ with physical temperature. The effective classical system now
resembles a hyper-slab with $d$ spatial dimensions (taken to be infinite
in extent) and
one temporal dimension of size $L_\tau \equiv \hbar \beta$. As phase
transitions depend sensitively upon the dimensionality of the system,
we expect the finiteness of $L_\tau$ to modify the critical
behavior, since at the longest length scales the system is now $d$ dimensional.

This modification can take two forms. First, it can destroy the transition
entirely so that the only critical point is at $T=0$. This happens in
the case of the 1D Josephson array. Its finite temperature physics
is that of an XY model on an infinite strip which, being a one dimensional
classical system with short-range forces,
is disordered at all finite values of $K$ (finite
temperatures in the classical language).

In the second form, the modification is such that
 the transition persists to $T \ne 0$ but crosses over
to a different universality class. For example, the problem of a 2D Josephson
junction array maps onto a 3($=$2+1) dimensional classical XY model.
Its phase diagram is illustrated in Fig.~(\ref{fig:phase_diagram}).
At $T=0$ the
QPT for the transition from superconductor to insulator
is characterized by the exponents of the 3D XY model. That is, it looks
just like the classical lambda transition in liquid helium with $K-K_{\rm
c}$ playing the role of $T-T_{\rm c}$ in the helium.
However at $T \ne 0$ the system is effectively two dimensional and
undergoes a transition of the 2D
Kosterlitz-Thouless\footnote{The Kosterlitz-Thouless phase transition
is a special transition exhibited by two-dimensional systems having
a continuous XY symmetry. It involves the unbinding of topological vortex
defects. See Goldenfeld (1992) and Chaikin and Lubensky (1995).}
 XY variety at a new,
smaller, value of $K$, much like a helium film. The Kosterlitz-Thouless
(KT) transition occurs on the solid line in Fig.~(\ref{fig:phase_diagram}).
We see that it is necessary to
reduce the quantum fluctuations (by making $K$ smaller) in order to allow
the system to order at finite temperatures. Above some critical
temperature, the system will become disordered (via the KT mechanism)
owing to thermal fluctuations. Of course, if we
make $K$ {\em larger} the quantum fluctuations are then so large that the
system does not order at any temperature, even zero. The region on the
$K$ axis (i.e., at $T=0$) to the right of the QCP in
Fig.~(\ref{fig:phase_diagram}) represents the quantum disordered
superconductor, that is, the insulator. At finite temperatures, no system
with a finite gap is ever truly insulating. However there is a crossover
regime, illustrated by the dotted line in Fig.~(\ref{fig:phase_diagram})
separating the regimes where the temperature is smaller and larger than the
gap.

At this point the reader might wonder how one learns anything at all
about the QPT if the effects of temperature are so dramatic. The
answer is that even though the finiteness of $L_\tau$ causes a {\em crossover}
away from the $T=0$ behavior, sufficiently close to the $T=0$ critical
point,
it does so in a fashion controlled by the physics at that critical point.
This is not an unfamiliar assertion. In the language of the renormalization
group, critical points are always unstable fixed points and lead to scaling
not because they decide where the system ends
up but because they control
how ``long'' it takes to get there. Here, instead of moving the system
away from the critical fixed point by tuning a parameter, we do so by
reducing its dimensionality.

Since the physics has to be continuous in temperature, the question
arises of how large the temperature has to be before the system
`knows' that its dimension has been reduced. %dropped `discontinuously' SLS
The answer to
this is illustrated in Fig.~(\ref{fig:finite-size}). When the coupling
$K$ is far away from the
zero-temperature critical coupling $K_{\rm c}$ the correlation length $\xi$
is not large and the corresponding correlation time $\xi_\tau \sim \xi^z$
is small. As long as the correlation time is smaller than the system
`thickness' $\hbar \beta$, the system does not realize that the temperature
is finite. That is, the characteristic fluctuation frequencies obey
$\hbar\omega \gg k_{\rm B}T$ and so are quantum mechanical in nature.
However as the critical coupling is approached, the correlation time grows
and eventually exceeds $\hbar\beta$. (More precisely, the correlation time
that the system
 would have had at zero temperature exceeds $\hbar\beta$; the
actual fluctuation correlation time is thus limited by temperature.)
At this point the system `knows' that
the temperature is finite and realizes that it is now effectively
a $d$-dimensional classical system rather than a $d+1$-dimensional system.

The formal theory of the effect of reduced dimensionality near
critical points, which quantifies the above considerations, is called
{\em finite size scaling} \cite{fss}.
For our problem
it asserts that for $[K-K_{\rm c}]/K_{\rm c} \ll 1$ and
$T \rightarrow 0$, physical quantities have the finite size scaling form,
\begin{equation}
{\cal O}(k, \omega, K, T) = L_\tau^{d_{\cal O}/z} {\cal O}(k L_\tau^{1/z},
\omega L_\tau , L_\tau/\xi_\tau).
\label{eq:finite_scaling}
\end{equation}
The interpretation of this form is the following. The quantity
$L_\tau\equiv\hbar\beta$ defined above leads, as discussed in more detail
below,
to a characteristic length $\sim L_\tau^{1/z}$
associated with the temperature. Hence the
prefactor $L_\tau^{d_{\cal O}/z}$ is the analog of the corresponding
prefactor in Eq.~(\ref{eq:homogen}). This same characteristic length is
the only one against which to measure the wave vector $k$. The associated
time
$L_\tau$ is the time scale against which to measure the frequency in the
second term. Finally the distance to the zero temperature critical
coupling is measured via the ratio of $L_\tau$ to the zero temperature
correlation time $\xi_\tau$.
The message here is that what matters is the ratio of the finite size
in the time direction
to the $T=0$ correlation length in that direction. We will return to
the uses of this form shortly.

Our considerations in this section also show us why the phase boundary in
Fig.~(\ref{fig:phase_diagram}) (solid line) and the crossover line (dashed
line) reach zero temperature in a singular way as the quantum critical
point is approached. Simple dimensional analysis tells us that the
characteristic energy scale ($\Delta$ for the insulator, $T_{\rm KT}$ for
the superfluid) vanishes near the critical point like $\hbar/\xi_\tau$,
implying
\begin{eqnarray}
\Delta &\sim& |K-K_{\rm c}|^{\nu z}\theta(K-K_{\rm c})\nonumber\\
T_{\rm KT} &\sim& |K-K_{\rm c}|^{\nu z}\theta(K_{\rm c}-K).
\end{eqnarray}

\subsection{The Quantum-Classical Crossover and the Dephasing Length}

We now turn to a somewhat different understanding and interpretation
of the effect of temperature that is conceptually of great importance.
Recall that the $T=0$ critical points of interest to us are
gapless and scale invariant, i.e., they have quantum fluctuations at
all frequencies down to zero. Temperature introduces a new energy scale
into the problem. Modes whose frequencies are larger than $k_B T/\hbar$
are largely unaffected,
while those with frequencies lower than $k_B T/\hbar$ become
occupied by many quanta with the consequence that they behave
{\em classically}. Put differently, the temperature cuts off
coherent quantum fluctuations in the infrared.

What we want to show next is that this existence of a quantum to classical
crossover frequency ($k_B T/\hbar$) leads to an associated length
scale for the same crossover, as alluded to in the previous section.
We shall refer to this length scale
as the dephasing length, $L_\phi$, associated with the QPT. The
temperature dependence of $L_\phi$ is easy enough to calculate. From
our imaginary time formalism we recall that quantum fluctuations are
fluctuations in the temporal direction.
Evidently these cannot be longer-ranged than the size of
the system in the time direction, $L_\tau = \hbar \beta$.
Since spatial and temporal correlations are linked via $\xi_\tau \sim \xi^z$,
it follows that the spatial correlations {\em linked} with quantum
fluctuations are not longer ranged than $L_\tau^{1/z}$. Since the spatial
range of quantum fluctuations is the dephasing length, we find
$L_\phi \sim 1/T^{1/z}$.

We use the term ``dephasing'' deliberately. Readers may know,
from other contexts where quantum effects are observed, that
such observation requires {\em phase coherence} for the relevant
degrees of freedom. In other words, interference terms should not be
wiped out by interactions with an ``environment'', i.e. other degrees of
freedom \cite{stern}. If dephasing takes place and the phase coherence
is lost, the resulting
behavior is classical. Thus our definition of $L_\phi$ is in line with
that notion. However, readers familiar with the notion of a dephasing length,
$\ell_\phi$, in mesoscopic condensed matter physics or the theory of
Anderson localization, might be concerned at our appropriation of this notion.
The concern arises because, in the standard lore in those fields, one is
often dealing with models of non-interacting or even
interacting electrons, whose quantum coherence is limited by degrees
of freedom, e.g. phonons, that are not being considered explicitly.
This has given rise to a tradition of thinking of $\ell_\phi$ as being
a length which is set {\em externally}.
Unfortunately this sort of separation of degrees of freedom should {\em not}
be expected to work near a QPT, since there one needs to keep track of
all {\em relevant} interactions. If a given interaction,
e.g. the Coulomb interaction, is relevant, then it already sits in the
Hamiltonian we need to solve and enters the calculation of $L_\phi$.
In contrast, if an interaction, e.g. coupling to phonons, is irrelevant
then we do not expect it to enter
the low energy physics as it should not in general affect the
quantum-classical crossover either.

Another way of formulating this is in terms of dephasing
{\em rates}. Since temperature is the only energy scale available,
a generic quantum critical point will be characterized by a
dephasing rate $\xi_\tau^{-1}$
that is linear in $T$, since we expect $\hbar/\xi_\tau \sim k_B T$. By
definition, irrelevant interactions, e.g. phonons, have an effective
coupling to the electrons which scales away to zero as one examines the
system on larger length and time scales. Hence such couplings will produce
a dephasing rate which
vanishes as $T^p$ with $p > 1$ and will therefore become negligible
compared to $T$ at low
temperatures.\footnote{Equivalently, the associated length scale will diverge
faster than $L_\phi$ as $T \rightarrow 0$ and hence will not control
the quantum-classical crossover of the relevant degrees of
freedom, since it is the shortest length that counts.
There are times when this
sort of reasoning can break down. These situations
involve operators that are irrelevant,
i.e., decrease in the infrared, but cannot be naively set to zero since that
produces extra singularities. Such operators are known in the trade as
``dangerous irrelevant operators'' and we will meet an example in the
next section, in the context of current scaling.}
%SMG696 Editor says he is lost here. What to do?

%SLS96 This is mystifying as it stands. Suggest rewrite and put in
% footnote.
%SMG96 is the following statement correct? I think so but we should check.
%For example, in dirty metals, the
%dephasing rate typically saturates at its maximum possible value
%$\sim k_{\rm B}T/\hbar$ as the metal-insulator transition is approached
%precisely because the Coulomb interactions control
%the fixed point. If in deriving Eq.(\ref{eq:branchcut}) we had considered
%the correlations at finite temperature (finite size in the time
%direction), we would have found that $k_{\rm B}T/\hbar$ now enters as a
%characteristic frequency in the imaginary part of the self-energy in the
%correlation function, indicating that the dephasing rate in this highly
%damped incoherent diffusive regime is indeed
%saturated at a value proportional to the temperature.

Thus we conclude that $L_\phi$ is the {\em unique} dephasing length
in the vicinity of a generic quantum critical point. Further discussion
and explicit examples
of the dissipative dynamics near a critical point can be found in the
article by Sachdev (1996).

\section{Experiments: QPTs in Quantum Hall Systems}

%%\noindent
%%{\em practical buzz-ideas: relevant perturbations, universal amplitudes}

We now turn from our somewhat abstract considerations to
examples of how one actually looks for a QPT in experiments. The
basic idea is relatively simple. We try to arrange that the system
is close to the conjectured critical point in parameter space ($K \sim
K_c$) and temperature ($T \sim 0$) and look for mutually consistent
evidence of scaling with various {\em relevant parameters}.
By these we mean either the deviation of the quantum coupling constant 
from its critical value $K-K_c$, %SMGnew
the temperature, or the wavevector,
frequency and amplitude of a probe. We call these relevant parameters
since when they are non-zero the response of the system has no
critical singularities due to the quantum critical point---hence the
analogy to relevant operators in renormalization 
group theory.\footnote{Consider for example a weak magnetic field applied
to a system undergoing  ferromagnetic ordering.  The magnetic field is
relevant and removes the sharp singularity in magnetization at the critical
temperature, replacing it with a rapid but smooth increase in magnetization.  
Likewise, measurements
at a non-zero frequency and wavevector do not exhibit singularities across
a transition.   For  quantum systems,
changes in the coupling and the temperature can produce more
subtle effects: they will cut off the critical fluctuations coming from the 
proximity to the quantum critical point, but either Goldstone modes coming from
a broken continuous symmetry or purely classical (thermal) fluctuations could
lead to independent singularities in the thermodynamics and response. We
saw an example of the latter in the persistence of a phase transition at finite
temperatures for the 2D Josephson junction array.} 
Additionally, we can look for 
universal critical amplitudes or amplitude ratios that are implicit in scaling 
forms for various quantities.

To see how this works, we will consider as a specific example,
a set of recent experiments on phase transitions in quantum Hall systems.
We derive various scaling relations appropriate to the
experiments, even though we do not actually know the correct theory
describing the QPT in this disordered, interacting fermi system. %SMG
%SMG696 NEED MORE BACKGROUND HERE
The very generality of the scaling arguments we will apply implies
that one need not have a detailed microscopic understanding of the physics
to extract useful information.
Nevertheless, we will start with some introductory
background for readers unfamiliar with the quantum Hall
effect \cite{bible1,bible2,bible3,bible4,bible5,bible6}.

The quantum Hall effect (QHE) is a property of a two dimensional electron
system placed in a strong transverse magnetic field, $B \sim 10T$. These systems 
are produced, using modern semiconductor fabrication techniques,
at the interface of two semiconductors with different band gaps. The
electronic motion perpendicular to the interface is confined to a potential
well $\sim 100$\AA\ thick. Because the well is so thin, the minimum
excitation energy perpendicular to the 2D plane ($\sim 200$K)
is much larger than the temperature ($\sim 1$K) and so motion in this third
dimension is frozen out, leaving the system dynamically two-dimensional.

As the ratio of the density to the magnetic field is varied at $T=0$,
the electrons condense into a remarkable sequence of distinct
thermodynamic phases.\footnote{The exact
membership of the sequence is sample specific but obeys certain selection
rules \cite{KLZ}.} These phases are most strikingly
characterized by their unique electrical transport properties, as
illustrated in Fig~(\ref{fig:qhdata}). Within each
phase the current flow is dissipationless, in that the longitudinal
resistivity, $\rho_{\rm L}$, that gives the electric field along
the direction of current flow ($E_{\rm L} = \rho_{\rm L} j$) vanishes.
At the same time that the longitudinal resistivity vanishes, the Hall
resistivity, $\rho_{\rm H}$, that gives the electric field transverse to
the direction of current flow ($E_{\rm H} = \rho_{\rm H} j$) becomes
quantized in rational multiples of the quantum of resistance
\begin{equation}
\rho_{\rm H} = \frac{h}{\nu_B e^2}
\label{eq:qhallr}
\end{equation}
where $\nu_B$ is an integer or simple rational fraction which serves to
distinguish between the different phases.
This quantization has been verified to an accuracy of about one part
in $10^7$ and to an even higher precision.

The QHE arises from a commensuration between electron density and magnetic
flux density, i.e. a sharp lowering of the energy when their ratio, the
filling factor $\nu_B$, takes on particular rational values. This
commensuration is equivalent to the existence of an energy gap in the
excitation spectrum at these ``magic'' filling factors and leads to
dissipationless flow at $T=0$ since degrading a small current requires
making arbitrarily small energy excitations which are unavailable. In the
absence of
disorder, Eqn.~(\ref{eq:qhallr}) follows straightforwardly, for example by
invoking Galilean invariance \cite{bible1}. As the magnetic field (or
density) is varied away from the magic filling factor, it is no longer
possible for the system to maintain commensuration over its entire area,
and it is forced to introduce a certain density of defects to take up
the discommensuration. In the presence of disorder, these defects, which
are the quasiparticles of the system, become localized and do not contribute
to the resistivities, which remain at their magic values. In this fashion,
we get a QH phase or a plateau.

Transitions between QH phases occur when too many quasiparticles have been
introduced into the original QH state and it becomes energetically
favorable to move to the ``basin of attraction'' of a different state
and its associated defects.
It might appear that these transitions between neighboring phases are
first order, since $\rho_{\rm H}$ jumps discontinuously
by a discrete amount between them, but they are not.
Qualitatively, they involve the quasiparticles of each phase which
are localized on a length scale, the localization length, that diverges
as the transition is approached from either side.
However, as these quasiparticles are
always localized at the longest length scale away from criticality, they
do not lead to dissipation ($\rho_{\rm L}=0$) and do not renormalize the
Hall resistivities of their respective phases.
Exactly at the transition they are delocalized and lead to a non-zero
$\rho_{\rm L}$. The shift in $\rho_{\rm H}$ on moving through the
transition can be understood in terms of either set of quasiparticles
condensing into a fluid state---there being an underlying duality in
this description.

In our description of the QH phases and phase transitions we have employed
a common language for all of them. We should note that this does not,
{\em ipso facto} imply that all quantum Hall transitions are in the same
universality class; however, experiments, as we discuss later, do seem to
suggest that conclusion. The reason for this caution is that different
QH states can arise from quite different physics at the microscopic level.
States with integer $\nu_B$ arise, to first approximation, from single
particle physics. An electron in a plane can occupy a {\em Landau level}
which comprises a set of degenerate states with energy
$(n+1/2) \hbar \omega_c$; these reflect the quantization of the classical
cyclotron motion having frequency $\omega_c = \frac{eB}{m}$ 
and the arbitrariness 
of the location of that motion in
the plane. When an integer number of Landau levels are full, and this
corresponds to an integer filling factor, excitations involve the promotion
of an electron across the cyclotron gap and we have the commensuration/gap
nexus necessary for the observation of the (integer) QHE. In contrast,
fractional
filling factors imply fractional occupations of the Landau levels,
with attendant macrosopic degeneracies, and they exhibit a gap only when
the Coulomb interaction between the electrons is taken into account (the
fractional QHE).\footnote{This leads to the remarkable feature that while
the quasiparticles of the integer states
are essentially electrons, those of the fractional states are
fractionally charged and obey fractional statistics.}
Readers interested in the details of this magic trick are encouraged
to peruse the literature.

Before proceeding to the details of experiments, we need to discuss two
important points about the units of the quantities measured in
electrical transport where two spatial dimensions are rather special.
Experiments measure resistances, which are ratios of total voltages
to current and these are related to local resistivities by ratios of
cross-sectional ``areas'' to lengths. In two dimensions, a cross-sectional
area {\em is} a length and consequently no factor of length intervenes
between the global and local quantities. In the QH phases, this has
the important implication that {\em no} geometrical factor affects the
measurement of the Hall resistance, which is why the ratio of fundamental
constants $h/e^2$ and hence the fine structure constant can be measured
to high accuracy on samples whose geometry is certainly not known to
an accuracy of one part in $10^7$.

What we have said above is a statement about the engineering dimensions of
resistance and resistivity. Remarkably, this also has an analog when it
comes to their {\em scaling} dimensions at a quantum critical point,
i.e. their scaling dimensions vanish.\footnote{See \cite{fgg,chaPRB}.
This is analogous to the behavior of the superfluid density at the
classical Kosterlitz-Thouless phase transition \cite{chaikin-lubensky}
and leads to a universal jump in it.}
Consequently, the resistivities vary as the zeroth power of the diverging
correlation length on
approaching the transition, i.e. will remain constant on either side.
Precisely at criticality they will be independent of the length scale used
to define them but can take values distinct from the neighboring
phases.

In this fashion, we have recovered from a purely scaling argument our
earlier conclusion that even though the quantum Hall transitions are
continuous, the resistivities at $T=0$ differ from the quantized values
{\em only} at critical points. Detecting the continuous transitions then
requires measurements at a non-zero temperature, frequency
or current, all of which lead to a more gradual
variation which can then be examined for scaling behavior.
Below are some examples of how that works.

\subsection{Temperature and Frequency Scaling}
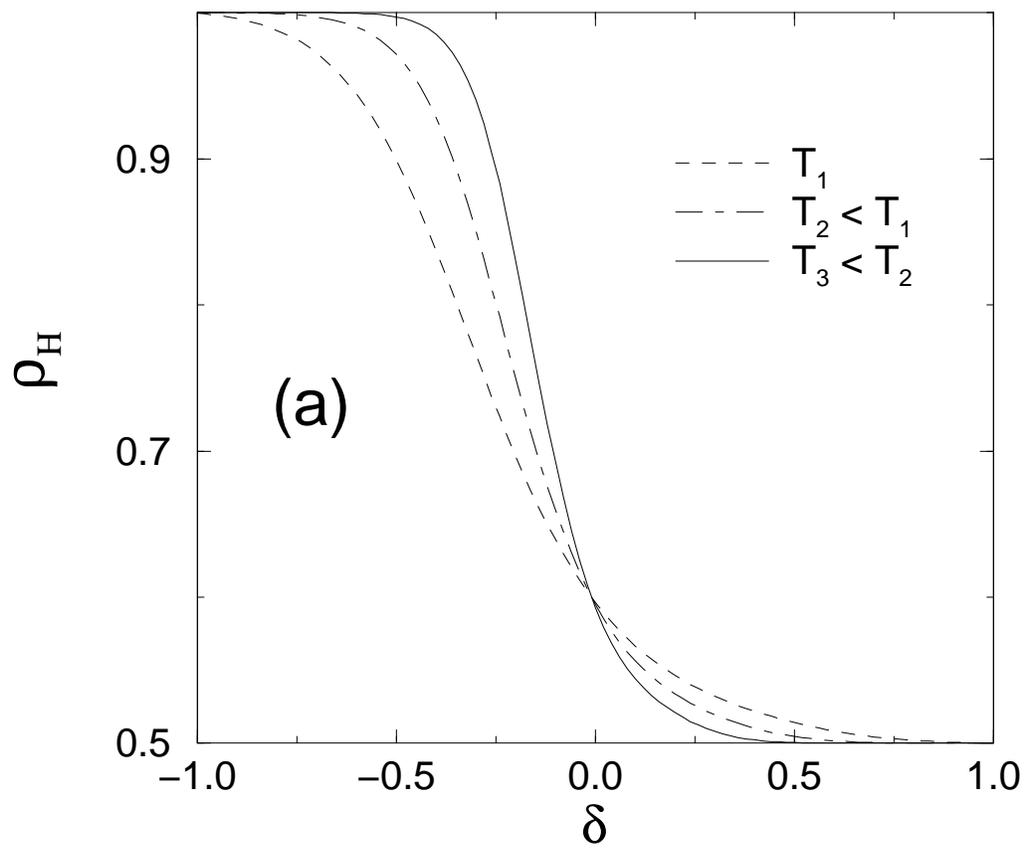
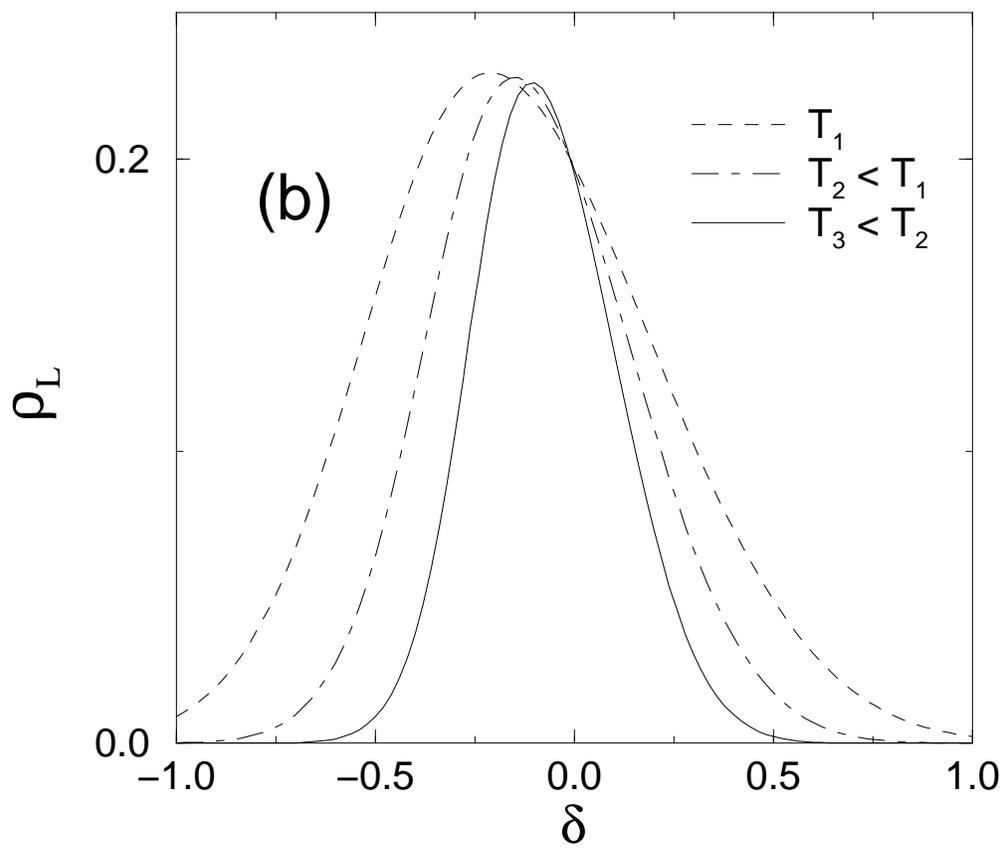
\begin{figure}
\caption[]{Caricature of the (ideal) magnetic field dependence of the Hall
(a) and longitudinal (b)
components of the resistivity tensor
near the quantum Hall critical point for the $\nu_B = 1 \longrightarrow 2$
transition, i.e. from the $\rho_{H} = h/e^2$ plateau
to the $\rho_{H} = h/2e^2$ plateau.
The parameter $\delta \equiv (B-B_{\rm c})/B_{\rm c}$ measures the
distance to the zero-temperature critical point. The three curves in each
panel correspond to three values of the temperature $T_1 > T_2 > T_3$.
At finite temperature the
variation of $\rho_{\rm L}$ and $\rho_{\rm H}$
is continuous, but sharpens up as the temperature is lowered.}
\label{fig:Tdata}
\end{figure}

Consider the caricature of a
typical set of data shown in Fig.~(\ref{fig:Tdata}).
Note that $\rho_{\rm H}$ interpolates
between its quantized values over a transition region of non-zero width,
while $\rho_{\rm L}$ is peaked in the same region, but is extremely small
outside the region. The change in the shape of these curves with temperature
can be understood on the basis of the finite size scaling form
\begin{equation}
\rho_{\rm {L/H}}(B,T,\omega)
= f_{\rm {L/H}}(\hbar\omega/ k_{\rm B}T,\delta/T^{1/\nu z}),
\label{eq:rhoscaling}
\end{equation}
where $\delta \equiv (B-B_{\rm c})/B_{\rm c}$ measures the distance to the
zero temperature critical point.
This form is equivalent to the general finite-size scaling form in
Eq.~(\ref{eq:finite_scaling}) except that we have assumed the limit $k=0$,
and used the previously cited result that the scaling dimension
of the resistivity vanishes in $d=2$ \cite{fgg,chaPRB}.
The first argument
in the scaling function here is the same as the second in
Eq.~(\ref{eq:finite_scaling}). The second argument in the scaling
function here is simply a power of the third argument in
Eq.~(\ref{eq:finite_scaling}). This change is inconsequential; it can be
simply absorbed into a redefinition of the scaling function.

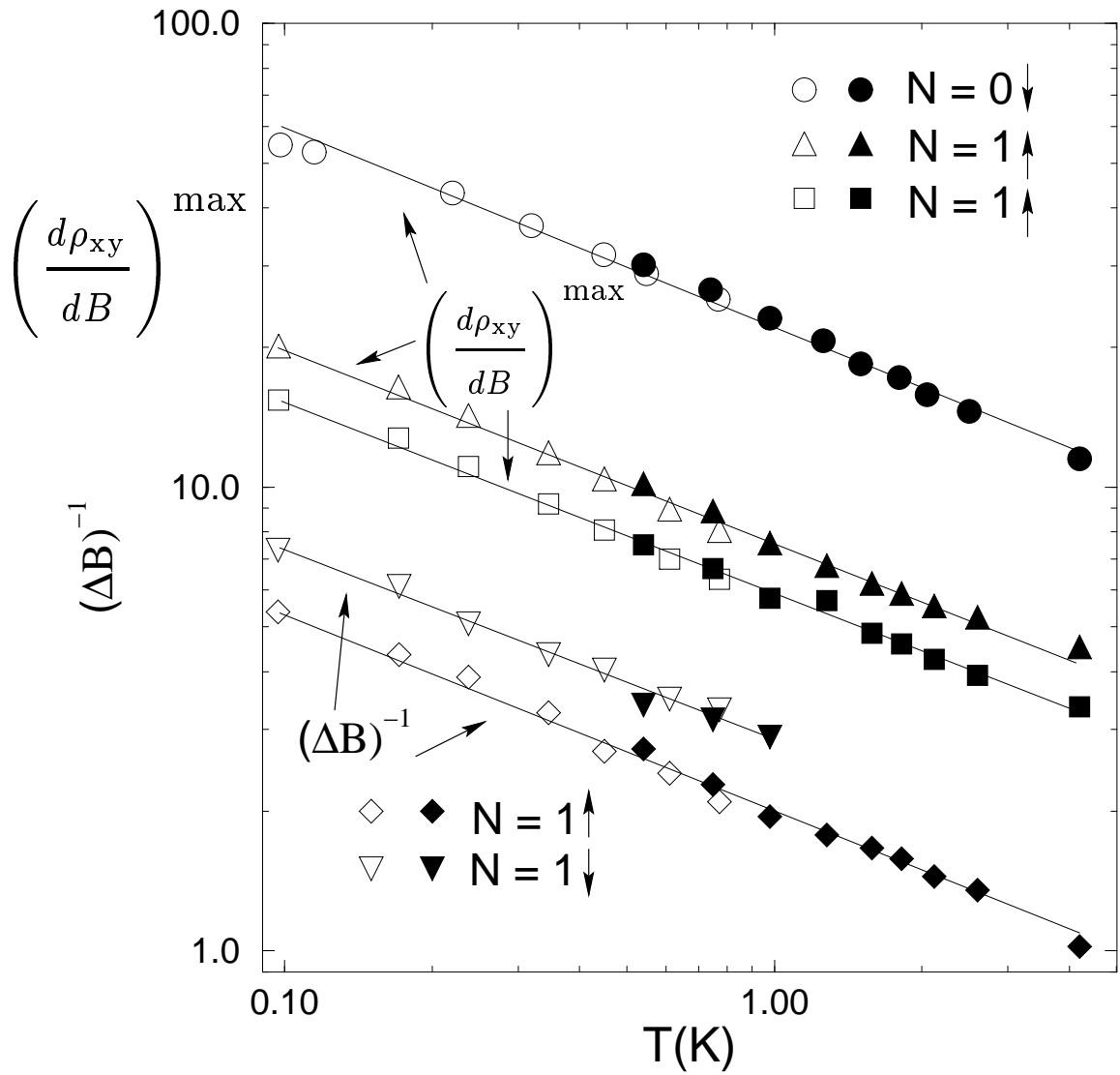
\begin{figure}
\caption[]{Data of Wei {\it et al.} (1988) showing the power-law
behavior of the width of the transition regions with temperature.
This double logarithmic plot shows two measures
of the inverse width, the maximum value of $\partial \rho_{H} /\partial B$
and the inverse of the field separation, $\Delta B$, between
$\partial \rho_{L}/\partial B$ maxima as a function of temperature. Both
are expected to scale as $T^{1/\nu z}$ and this data yields $1/\nu z =
0.42 \pm 0.04$. The labels for the symbols refer to different transitions:
$N = 0 \downarrow$ to the $\nu_B = 1 \longrightarrow 2$ transition,
$N = 1 \uparrow$ to the $\nu_B = 2 \longrightarrow 3$ transition
and $N = 1 \downarrow$ to the $\nu_B = 3 \longrightarrow 4$
transition. The units of $\Delta B$ are Tesla (T) and the units of
$\partial \rho_{H} /\partial B$ are $10^3\Omega/{\rm T}$}.
%Figure 2 of their PRL.
\label{fig:wei-data}
\end{figure}

%SLS
First, let us consider a DC or $\omega =0$ measurement. In this case
our scaling form implies that the resistivities are not independent
functions of $\delta$ (or $B$) and $T$ but instead are functions
of the single
{\em scaling variable} $\delta/T^{1/\nu z}$. Hence the effect of lowering
$T$ is to rescale the deviation of the field from its critical
value by the factor $T^{1/\nu z}$. It follows that the
transition appears sharper and sharper as the temperature is lowered,
its width vanishing as a universal power of the temperature,
$\Delta B \sim T^{1/\nu z}$.
In Fig.~(\ref{fig:wei-data}) we show the pioneering data of Wei {\it et al.}
(1988) that indeed shows such an algebraic dependence for several
different
transitions all of which yield the value $1/\nu z \approx 0.42$. These
transitions are between integer quantum Hall states.
Remarkably, this temperature scaling behavior seems to be ubiquitous
at quantum Hall transitions and suggests that there is a single underlying
fixed point for all of them. It was shown by Engel {\it et al.} (1990)
that it holds at transitions between two fractional quantum Hall states.
Subsequently,
Wong {\it et al.} (1995) found the same scaling for the transition
between a Hall state and the insulator. In Fig.~(\ref{fig:shahar-data}) we
show some recent data
%\cite{shahar-thesis}
%SMG had to do this to get citation in before
% references ended. Previously only appeared in figure caption which
% is after the references ended. TeX got confused.
of Shahar (1995) near another such transition, plotted both
as a function of the magnetic
field at several values of $T$, and against the scaling variable
$\delta/T^{1/\nu z}$, exhibiting the data collapse characteristic
of the critical region.

Consider now the results of measurements at non-zero frequencies. In
their full generality these require a two variable scaling
analysis \cite{engel-two} but we focus instead on two distinct
regimes. In the regime
$\hbar \omega \ll k_B T$ we expect that the behavior of the scaling
function will be governed by its $\omega=0$ limit analyzed previously,
i.e. at small $\omega$ we expect the scaling to be dominated
by $T$. In the second regime, $\hbar \omega \gg k_B T$, we expect
the scaling to be dominated by $\omega$ and the scaling function to
be independent of $T$. In order for the temperature to drop out,
the scaling function in Eq.~(\ref{eq:rhoscaling}) must have the form
\begin{equation}
f(x,y) \sim \tilde{f} (y x^{-1/\nu z})
\end{equation}
for large $x$ so that the scaling variables
conspire to appear in the combination,
\begin{equation}
\left(\frac{\hbar \omega}{k_B T}\right)^{-1 /\nu z} {\delta \over T^{1/\nu z}}
\sim {\delta \over \omega^{1/\nu z}} \ .
\end{equation}
It follows that at high frequencies the resistivities are functions
of the scaling variable $\delta/\omega^{1/\nu z}$ and that the width
of the transition regions scales as $\omega^{1/\nu z}$.
Fig.~(\ref{fig:engel-data}) shows frequency dependent
conductivity\footnote{The conductivities scale in exactly the same fashion
as the resistivities.} data of Engel {\it et al.} (1993) which exhibits
this algebraic increase in the width of the transition region with
frequency and yields a value of $\nu z$ consistent with the temperature
scaling.

We should note an important point here. As the ratio $\hbar \omega/k_B T$
is varied in a given experiment we expect to see a crossover between
the $T$ and $\omega$ dominated scaling regimes. The criterion for
this crossover is $\hbar \omega \approx k_B T$.
The observation by Engel {\it et al.} (1990), that this is indeed the
correct crossover criterion (see Fig.~(\ref{fig:engel-data})) is
important for two reasons. First, it involves $\hbar$ and clearly implies
that {\em quantum} effects are at issue. Second it implies that $T$ is the
relevant infrared scale. If dephasing effects coming from coupling to
some irrelevant degree of freedom
were important, one would expect the crossover to take place
when $\omega \tau \approx 1$, where $1/\tau$ is some microscopic
scattering or relaxation rate associated with this coupling.
Since the coupling is irrelevant it will, as noted earlier,
give a scattering rate that vanishes as $AT^p$ where $p$ is greater
than unity and $A$ is non-universal \cite{sls-stevek} (e.g., it depends on
the precise value of the electron-phonon coupling constant for the material).
In contrast, what is
observed is that the relaxation rate obeys $1/\tau = Ck_{\rm B}T/\hbar$
where $C$ is a {\em universal} \cite{sachdevIUPAP}
dimensionless constant of order unity.
%SMG696 UNIVERSAL IS OK???

It is important to note that frequency scaling does not give
us any new information on exponents that we did not already have from the
temperature scaling. The main import of frequency scaling is its ability
to confirm the quantum critical nature of the transition by showing that
the characteristic time scales have diverged, leaving the temperature
itself as the only frequency scale.
%SMG696 EDITOR SAYS TOO FAST HERE

\subsection{Current Scaling}
A third relevant parameter that is experimentally useful is the
magnitude of the measuring current or, equivalently, of the applied
electric field. In talking about resistivities we have assumed that there
is an ohmic regime at small currents, i.e., a regime in which the voltages
are linear in the current.
%%%If the measuring current is increased beyond
%%%this regime, non-linearities set in.
In general, there is no reason to
believe that the non-linear response can be related to equilibrium
properties---i.e., there is no fluctuation-dissipation theorem beyond
the linear regime. However, in the vicinity of a critical point
we expect the dominant non-linearities to come from critical
fluctuations. At $T=0$, the electric field scale for these can be estimated
from an essentially dimensional argument. We start by defining a
characteristic length $\ell_E$ associated with the electric field. Imagine
that the system is at the critical point so that $\ell_E$ is the only
length scale available. Then the only characteristic time for
fluctuations of the system will scale like $\ell_E^{+z}$.
We can relate the length $\ell_E$ to the electric field that produces it by
\begin{equation}
eE\ell_E \sim \hbar\ell_E^{-z}.
\end{equation}
This expression simply equates the energy gained from the electric field by
an electron moving a distance $\ell_E$ to the characteristic energy
of the equilibrium system at that same scale. Thus
\begin{equation}
\ell_E \sim E^{-1/(1+z)}.
\end{equation}
If the system is not precisely at the critical point, then it is this
length $\ell_E$ that we should compare to the correlation length
\begin{equation}
\frac{\ell_E}{\xi} \sim \delta^{\nu} E^{-1/(1+z)}\sim
\left(\frac{\delta}{E^{1/\nu(z+1)}}\right)^\nu.
\end{equation}
From this we find that the
non-linear DC resistivities for a 2D system obey the scaling forms
\begin{equation} \label{eq:iscaling}
\rho_{L/H}(B,T,E) = g_{\rm L/H}(\delta/T^{1/\nu z}, \delta/E^{1/\nu(z+1)}) .
\label{current_scaling1}
\end{equation}
This is a very useful result because it tells us that electric field
scaling will give us new information not available from temperature scaling
alone. From temperature scaling we can measure the combination of
exponents $\nu z$. Because an electric field requires multiplication by
one power of the correlation
length to convert it to a temperature (energy),
electric field scaling measures the
combination of exponents $\nu(z+1)$. Thus the two measurements can be
combined to separately determine $\nu$ and $z$.
%Again there are two scaling regimes. At small currents the scaling is
%dominated by $T$ and we find $I$ independent values of $\rho_{L/H}$.
%At fields larger than the crossover scale $E_0(T) \sim T^{(1+z)/z}$, the
%scaling is dominated by the current. Consequently the resistivities
%become functions of the scaling variable $\delta/I^{1/\nu(z+1)}$ and
%the width of the transition region scales as $\Delta B \sim I^{1/\nu(z+1)}$.
The data of Wei {\it et al.} (1994), Fig.~(\ref{fig:wei-Idata}),
confirm this
%%%bear out the power law scaling of the width
and yield the value $\nu(z+1) \approx 4.6$.
Together the $T$, $\omega$ and $I$ scaling experiments lead to the
assignment $\nu \approx 2.3$ and $z \approx 1$.

Equation~(\ref{current_scaling1}) tells us that there are two scaling
regimes. At sufficiently high temperatures, $L_\phi \ll \ell_E$ and
the scaling is controlled by the
temperature. Below a crossover temperature scale
\begin{equation}
T_0(E) \sim \ell_E^{-z} \sim E^{z/(1+z)},
\label{eq:T_0(E)}
\end{equation}
$L_\phi > \ell_E$ and
the scaling is controlled by the electric field $E$ (or equivalently, the
applied current $I$). One
might be tempted to identify $T_0(E)$ as the effective temperature of the
electrons in the presence of the electric field, but this is not strictly
appropriate since the system is assumed to have been driven out of
equilibrium on length scales larger than $\ell_E$.

This quantum critical scaling picture explicitly assumes that the slow internal
time scales of the system near its critical point control the response to
the electric field and implicitly assumes that we can
ignore the time scale which determines how fast the Joule heat
can be removed by phonon radiation.
Thus this picture is quite distinct from that of a simple heating scenario
in which the electron gas itself equilibrates rapidly, but undergoes
a rise in temperature if there is a bottleneck for the energy deposited by
Joule heating to be carried away by the phonons.
This effect can give rise to an apparent
non-linear response that is, in fact, the linear response of the electron
gas at the higher temperature. The power radiated into phonons at low
electron temperatures scales as
\begin{equation}
P_{\rm ph} = A T_{\rm e}^\theta,
\end{equation}
where $\theta = 4-7$ depending on details \cite{chow}.
Equating this to the Joule
heating (assuming a scale invariant conductivity) yields an electronic
temperature
\begin{equation}
T_{\rm elec} \sim E^{2/\theta}.
\label{eq:T_e(E)}
\end{equation}
We now have a paradox. The more irrelevant phonons are at low temperatures
(i.e., the larger $\theta$ is), the smaller is the exponent $2/\theta$
and hence {\em the more singular is the temperature rise
produced by the Joule heat}. Comparing Eqs.(\ref{eq:T_0(E)})
and (\ref{eq:T_e(E)})
we see that for
\begin{equation}
\frac{2}{\theta} < \frac{z}{z+1},
\end{equation}
we have
\begin{equation}
T_{\rm elec}  > T_0(E).
\end{equation}
That is, we then have that
the temperature rise needed to radiate away sufficient power
is larger than the characteristic energy (`temperature')
scale predicted by the quantum critical scaling picture.
In this case the phonons are `dangerously
irrelevant' and the simple quantum critical scaling prediction fails.
It happens that for the case of GaAs, which is piezoelectric, $2/\theta =
1/2$ which gives the same singularity exponent as the quantum critical model
 $z/(z+1) = 1/2$ (since $z=1$).
Hence both quantum critical and heating effects are important.
(The phonon coupling is `marginally dangerous'.)
This result is discussed in more detail
 elsewhere \cite{chow,girvin-sondhi-fisher}.

\subsection{Universal Resistivities}

The final signatures of critical behavior which we wish to discuss
are universal amplitudes, and, more generally, amplitude ratios.
These are readily illustrated in the quantum Hall problem without
considering their general setting, for which we direct the reader
to the literature
\cite{amplitudes1,amplitudes2,amplitudes3,2dafm2,sachdevIUPAP}.
Note that the scaling forms (\ref{eq:rhoscaling}) and
(\ref{eq:iscaling}) imply that the resistivities at $B=B_c$ in the
critical region are independent of $T, \omega$ and $I$. Under certain
assumptions it is possible to argue that they are, in fact, universal
\cite{KLZ,fgg}. The observation of such universality
between microscopically different samples would then be strong evidence
for an underlying QPT as well.

Recently Shahar {\it et al.} (1995) have carried out a study of the
critical resistivities at the transition from the $\nu_B =1$ and $1/3$
quantum Hall states to the insulating state. An
example of their data is shown in Fig.~(\ref{fig:shahar-data}). Notice
that there exists a critical value of $B$ field at which the resistivity
is temperature-independent. For $B<B_{\rm c}$ the resistivity scales
upward with decreasing $T$, while for $B>B_{\rm c}$, it scales downward
with decreasing $T$. Since we can think of lowering $T$ as increasing the
characteristic length scale $\L_\phi$ at which we examine the system, we
see that the point where all the curves cross is the scale-invariant point
of the system and hence must be the critical point.

Shahar {\it et al.} (1995) find that at these critical points,
$\rho_{\rm L}$ is {\em independent} of the sample studied and in fact
appears to be $h/e^2$ within experimental error for both transitions.
Preliminary studies \cite{shahar_privcomm} also seem to find
sample-independent values of $\rho_{\rm H}$ at the critical points with
values of $h/e^2$ and $3h/e^2$ for the two transitions.

%SLS
\subsection{Unresolved Issues}

As we have tried to indicate, the success of experimental work
in making a case for universal critical behavior at
transitions in the quantum Hall regime is impressive.
However, not everything is settled on this score. Apart from the
delicate issues surrounding the interpretation of the current
scaling data mentioned earlier, there is one significant puzzle.
This concerns the failure of $\rho_L$ at the transition between two
generic QH states to exhibit a $T$-independent value at a critical field
even as the width of the curve exhibits algebraic behavior.\footnote{Hence
our unwillingness to plot the actual traces in Fig.~(\ref{fig:Tdata}).}
This is generally believed to stem from macroscopic inhomogeneities
in the density and some recent theoretical work offers support
for this notion \cite{ruzinetc}. Nevertheless, this is an issue that 
will need further work. The transitions to the insulator studied more
recently, are believed to be much less sensitive to this problem, and 
hence the consistency of the data on those is encouraging. However, in
these cases the temperature range over which there is evidence for 
quantum critical scaling is quite small as in the data in 
Fig.~(\ref{fig:shahar-data}) which leads us to a general caveat.

Evidence for power laws and scaling should properly consist of overlapping
data that cover several decades in the parameters. The various
power law dependences that we have exhibited span at best two decades,
most of them fewer and the evidence for data collapse within the error bars of
the data exists only over a small range of the scaled variables. Consequently,
though the overall picture of the different types of
data is highly suggestive, it cannot really
be said that it does more than indicate consistency with the scaling expected
near a quantum critical point. Regrettably, there is at present no example
of a quantum critical phase transition as clean as
the remarkable case of the classical lambda transition in superfluid
helium for which superb scaling can be demonstrated. \cite{ahlers} 

On the theoretical front the news is mixed.
Remarkably, the experimental value of the correlation length exponent
$\nu \approx 2.3$ is consistent with numerical
calculations of the behavior of {\em non-interacting} electrons in a strong
magnetic field \cite{bodo}. Also, the critical resistivities at the
transition from
the $\nu_B=1$ state to the insulator are also consistent with these
calculations \cite{bodo}. This agreement is still a puzzle at this time,
especially
as the value of the dynamic scaling exponent $z \approx 1$ strongly
suggests that Coulomb interactions are playing an important role.
The evidence for a super-universality of the transitions, however
does have some theoretical support in the form of a set of physically
appealing ``correspondence rules'' \cite{KLZ}. Unfortunately, their
{\em a priori} validity in the critical regions is still unclear.
\cite{leeandwang}
In sum, theorists have their work cut out for them!

\section{Concluding Remarks, Other Systems}

%%\noindent
%%{\em valedictory buzz}

Let us briefly recapitulate our main themes. Zero temperature phase transitions
in quantum systems are fundamentally different from finite temperature
transitions in classical systems in that their thermodynamics and dynamics
are inextricably mixed. Nevertheless, by means of the path integral
formulation of quantum mechanics, one can view the statistical mechanics of a
d-dimensional $T=0$ quantum system as the statistical mechanics of a
d+1 dimensional classical system with a fake temperature which
 is some measure of zero-point
fluctuations in the quantum system. This allows one to apply ideas and
intuition developed for classical critical phenomena to quantum critical
phenomena. In particular this leads to an understanding of the $T \ne 0$
behavior of the quantum system in terms of finite size scaling
and to the identification of a $T$-dependent length scale, $L_\phi$, that
governs the crossover between quantum and classical fluctuations.
%SLS
The identification of QPTs in experiments relies upon finding
scaling behavior with relevant parameters. These are the temperature
itself and the frequency, wavelength and amplitude of various probes.
Additional signatures are universal values of certain dimensionless
critical amplitudes such as the special case of
resistivities at critical points in conducting systems in d=2
and, more generally, amplitude ratios.

In this Colloquium we have illustrated these ideas in the context of
a single system, the two dimensional electron gas in the quantum Hall
regime. The ideas themselves are much more widely applicable. Interested
readers may wish to delve, for example, into work on the one dimensional
electron gas \cite{luther,emery}, metal insulator
transitions in zero magnetic field (``Anderson-Mott transitions'')
\cite{bk}, superconductor-insulator
transitions
\cite{wallinetalPRB,chaPRB,sit1,sit2,sit3,sit4,sit5,sit6,sit7,sit8,%
sit9,sit10,sit11,sit12,sit13,sit14},
two-dimensional antiferromagnets associated with high temperature
superconductivity
\cite{sachdevIUPAP,2dafm6,2dafm1,2dafm2,2dafm3,2dafm4,2dafm5} and
magnetic transitions in metals \cite{hertz,fm1,fm2,fm3}. This list is by
no means exhaustive and we are confident that it will continue to expand
for some time to come!

\acknowledgements

It is a pleasure to thank R. N. Bhatt, M. P. A. Fisher, E. H. Fradkin,
M. P. Gelfand, S. A. Kivelson, D. C. Tsui and H. P. Wei for numerous
helpful conversations. We are particularly grateful to K. A. Moler,
D. Belitz, S. Nagel, T. Witten, T. Rosenbaum, and U. Zuelicke
 for comments on early versions of the manuscript.
DS is supported by the NSF, the work at Indiana
is supported by NSF grants DMR-9416906, DMR-9423088 and DOE grant 
DE-FG02-90ER45427, and SLS is supported
by the NSF through DMR-9632690 and by the A. P. Sloan Foundation.

\newpage
\appendix
\renewcommand{\theequation}{\Alph{section}.\arabic{equation}}
\section{}

In this Appendix we briefly outline the derivation \cite{wallinetalPRB} 
of the expression for the matrix elements 
\begin{equation}
M \equiv \langle \{\theta(\tau_{j+1})\} |
e^{-\frac{\delta\tau}{\hbar}H} |\{\theta(\tau_j)\}\rangle
\end{equation}
appearing in Eq.~(\ref{eq:statwgt}).  The hamiltonian contains a `kinetic
energy'
\begin{equation}
T = \frac{C}{2} \sum_j V_j^2 = \frac{E_{\mathrm C}}{2}
\sum_j\left(-i\frac{\partial}{\partial\theta_j}\right)^2,
\end{equation}
where $E_{\mathrm C} \equiv \frac{(2e)^2}{C}$, and a `potential energy'
\begin{equation}
V \equiv - E_{\rm J} \cos\left(\hat\theta_j - \hat\theta_{j+1}\right).
\end{equation}
For sufficiently small $\delta\tau$ we can make the approximation
\begin{equation}
e^{-\frac{\delta\tau}{\hbar}
H} \approx
e^{-\frac{\delta\tau }{\hbar} T}e^{-\frac{\delta\tau}{\hbar} V}.
\end{equation}
%%%%%%%%%%which is valid to order $(\delta\tau)^2$.

Inserting a complete set of angular momentum eigenstates $|\{m_k\}\rangle$
(defined for a single site by 
$\langle\theta_k|m_k\rangle = e^{im_k\theta_k}$)
yields
\begin{equation}
M = \sum_{\{m\}}\langle \{\theta(\tau_{j+1})\} |
e^{-\frac{\delta\tau}{\hbar} T} 
|\{m_k\}\rangle
\langle \{m_k\}|
e^{-\frac{\delta\tau}{\hbar}V} 
|\{\theta(\tau_j)\}\rangle.
\end{equation}
We can now take advantage of the fact that $V$ is diagonal in the angle
basis and $T$ is diagonal in the angular momentum basis to obtain
\begin{equation}
M = \sum_{\{m\}} 
e^{-\frac{\delta\tau}{2\hbar}E_{\mathrm C} \sum_k m_k^2}
\,e^{i m_k[\theta_k(\tau_{j+1}) - \theta_k(\tau_j)]}\,\,
e^{+\frac{\delta\tau }{\hbar} E_{\mathrm J}
\sum_k\cos\left[\theta_k(\tau_{j+1})-\theta_k(\tau_j)\right]}
\label{eq:M}
\end{equation}

Because $\delta\tau$ is small, the sum over the $\{m\}$ is slowly
convergent.  We may remedy this by using the Poisson summation formula
\cite{wallinetalPRB}
\begin{equation}
\sum_m e^{-\frac{\delta\tau}{2\hbar}E_{\mathrm C}m^2} e^{im\theta}
= \sqrt{\frac{\pi\hbar}{\delta\tau E_{\mathrm C}}}\sum_n
e^{-\frac{\hbar}{2E_{\mathrm C}\delta\tau}(\theta + 2\pi n)^2}.
\end{equation}
This periodic sequence of very narrow gaussians is 
(up to an irrelevant constant prefactor)
the Villain approximation to 
\begin{equation}
e^{+\frac{\hbar}{E_{\mathrm C}\delta\tau}
\cos(\theta)}.
\end{equation}
Strictly speaking, we should keep $\delta\tau$ infinitesimal.  However
we may set it equal to the natural ultraviolet cutoff, the inverse
of the Josephson plasma frequency
$\delta\tau = \hbar/\sqrt{E_{\mathrm C}{E_J}}$, without changing the
universality class.
Substituting this result into Eq.(\ref{eq:M}) yields Eq.(\ref{eq:equivXY}) with
the same coupling constant $K = \sqrt{E_{\mathrm C}/E_{\mathrm J}}$ in both
the space and time directions.

%SLS
\begin{table}
\caption{Analogies}
\begin{tabular}{cc} % In second brace, l = left, r = right,
 % c = centered and d = decimal justification.
Quantum & Classical \\

$d$ space, $1$ time dimensions & $d+1$ space dimensions \\

Coupling constant $K$ & Temperature $T$ \\

Inverse temperature $\beta$ & Finite size $L_\tau$ in `time' direction \\

Correlation length $\xi$ & Correlation length $\xi$ \\

Inverse Characteristic Energy $\hbar/\Delta, \hbar/k_{\rm B}T_{\rm c}$
& Correlation length in the `time' direction $\xi_\tau$ \\
%\tableline % Creates a horizontal line.
\end{tabular}
\label{tableA}
\end{table}

\begin{figure}
\bigskip
\caption[]{Schematic representation of a 1D Josephson junction array. The
crosses represent the tunnel junctions between superconducting segments
and $\theta_i$ are the phases of the superconducting order parameter in
the latter.}
\label{fig:JJdiagram}
\end{figure}

\begin{figure}
%\centerline{\psfig{figure=JJpath.xfig.eps,height=2.0in,width=2.0in}}
\bigskip
\caption[]{Typical path or time history of a 1D Josephson junction array.
Note that this is equivalent to one of the configurations of
a 1+1D classical XY model. The long range correlations shown here
are typical of the superconducting phase of the 1D array or equivalently,
of the ordered phase of the classical model.}
\label{fig:JJpath}
\end{figure}

\begin{figure}
%\centerline{\psfig{figure=JJpath2.xfig.eps,height=2.0in,width=2.0in}}
\bigskip
\caption[]{Typical path or time history of a 1D Josephson junction array
in the insulating phase where correlations fall off exponentially
in both space and time. This corresponds to the disordered phase
in the classical model.}
\label{fig:JJpath2}
\end{figure}

\begin{figure}
\caption[]{Illustration of discrete space-time lattices with the same
coupling K. The lattice constant in the time direction is the same
$\delta\tau$ in both cases, but the corresponding physical temperature
(determined from $\hbar\beta = L_\tau \delta\tau$) for the lattice
in the lower panel is half that of the lattice in the upper panel.}
\label{fig:two-sizes}
\end{figure}

\begin{figure}
\caption[]{Illustration of the phase diagram for a Josephson junction array
in two dimensions. $K$ is the quantum fluctuation parameter and $T$ is the
physical temperature. The solid line represents the Kosterlitz-Thouless
critical temperature for the phase transition from the normal state
to the superfluid. For $K$ greater than its critical value, the system is
insulating at zero temperature. For any finite temperature it is not
insulating. The dashed line represents the crossover from temperatures
smaller than the ($T=0$ insulating) gap to temperatures greater than the gap.
This is not a true phase transition, however the conductivity can be
expected to increase rapidly as the temperature goes above this line.
}
\label{fig:phase_diagram}
\end{figure}

\begin{figure}
%\centerline{\psfig{figure=fig:finite-size.xfig.eps,height=2.0in,width=2.0in}}
\bigskip
\caption[]{Illustration of the growing correlation volume as the (T=0)
critical coupling $K_{\rm c}$ is approached in a system with finite extent
in the temporal direction. In (a) the correlation time is much shorter
than $\hbar\beta$. In (b) it is comparable. In (c) the system is very
close to the critical point and the correlation time
(that the system would have had at zero temperature)
greatly exceeds $\hbar\beta$.
Once $\hbar\beta < \xi_\tau \sim \xi^z$, the
system realizes that it is effectively $d$-dimensional and not
$d+1$-dimensional. The actual correlation time saturates at $\hbar \beta$
and the corresponding $T=0$ correlation length at which this occurs
is the quantum-to-classical crossover length.}
\label{fig:finite-size}
\end{figure}

\begin{figure}
\caption[]{Transport data in the quantum Hall regime \cite{tsui}.
The diagonal trace
is $\rho_H$ while the oscillating trace is $\rho_L$ for a sample with
density $3 \times 10^{11}/{\rm cm}^2$ and mobility $1.3 \times 10^{11}
{\rm cm}^2/{\rm Vs}$ at $T=85 mK$. Note that the latter tends to vanish
when the former has a plateaux. The arrows mark various filling factors
while the $N$ values at the top mark the Landau level occupied by the
electrons.}
\label{fig:qhdata}
\end{figure}

\begin{figure}
\caption[]{$\rho_{L}$ data of Shahar {\em et al.} (1995)
near the $\nu_B=1/3$ to insulator transition. a) The raw data showing
a critical magnetic field with a $T$-independent
$\rho_{L}^c \approx h/e^2$. b) Double logarithmic scaling plot of the
the same data. Note the systematic improvement in the collapse close
to the critical point. [The data points at the far right lie outside the
scaling regime.]
The apparent discrepancy between the value
$1/\nu z = 0.48$ obtained here and the earlier reported
values is not significant.  Reliable error estimates for critical 
exponents are notoriously hard to obtain.}
%After Fig 4.9 of Shahar's thesis.
\label{fig:shahar-data}
\end{figure}

\begin{figure}
\caption[]{Data of Engel {\it et al.} (1990) showing the power-law
behavior of the width of the transition regions with the frequency of the
measurement. (a) double logarithmic plot showing the width,
$\Delta B$, defined as the field separation between $\partial
\sigma_{L}/\partial B$ maxima, at a fixed temperature of 50 mK, as a
function of frequency at five different transitions:
$\nu_B = 1 \longrightarrow 2$ (crosses),
$\nu_B = 2 \longrightarrow 3$ (open circles),
$\nu_B = 3 \longrightarrow 4$ (filled circles),
$\nu_B = 4 \longrightarrow 6$ (open triangles) and
$\nu_B = 6 \longrightarrow 8$ (filled triangles). The
last two are believed to be two transitions too closely spaced to
be resolved at accessible temperatures, and show anomalous scaling
as a consequence.
(b) data for the $\nu_B = 1 \longrightarrow 2$
transition at three different temperatures. Note that
$\Delta B$ is independent of frequency below a crossover value roughly
equal to the temperature (1 GHz is roughly 50mK), and above the crossover,
scales as $f^{1/\nu z}$ with $1/\nu z = 0.43$ consistent with the value
obtained from the temperature scaling.}
% Figure 4b of their PRL.
\label{fig:engel-data}
\end{figure}

\begin{figure}
\caption[]{Data of Wei {\it et al.} (1994) showing power-law behavior of
the effective electron ``temperature'' $T_e$ with current $I$
at two different transitions: $\nu_B = 5 \longrightarrow 6$
(closed symbols) and $\nu_B = 1/3 \longrightarrow 2/5$ (open
symbols, reduced by a factor of 6).
$T_e$ is {\em experimentally}
defined as the equilibrium temperature at which the maximum of the slope 
$\partial \rho_h/\partial B$ in {\em linear} response equals the 
measured maximum slope of the same quantity in non-linear response at low
bath temperatures.  In the quantum critical model, $T_e$ is not
interpreted as a true temperature, but
rather as the disequilibrium energy scale $T_0(E)$ in Eq.(\ref{eq:T_0(E)}).
In the heating model, $T_e$ is interpreted as the quasi-equilibrium
electronic temperature $T_{\mathrm elec}$ in Eq.(\ref{eq:T_e(E)}).
The temperature labels in units of mK at the left hand side
represent the bath temperature at which the current dependent
transport is measured. The dashed reference line under each data set
has a slope of 0.5, and the solid line 0.4. Assuming $T_e \sim T_0$ (see
text) where $T_0 \sim I^{z/(1+z)}$, the data suggest the assignment $z=1$.
Analysis in terms of the heating model yields the exponent $\theta = 4$
(see text).
Inset: $\rho_H$ vs.~$B$ at $T = 100$\,mK for
three different $I$'s in the FQHE for $1/3 < \nu < 2/5$. This
illustrates how the transition between Hall plateaus
sharpens up as the current is reduced, much as it
does when the temperature is reduced.}
\label{fig:wei-Idata}
\end{figure}

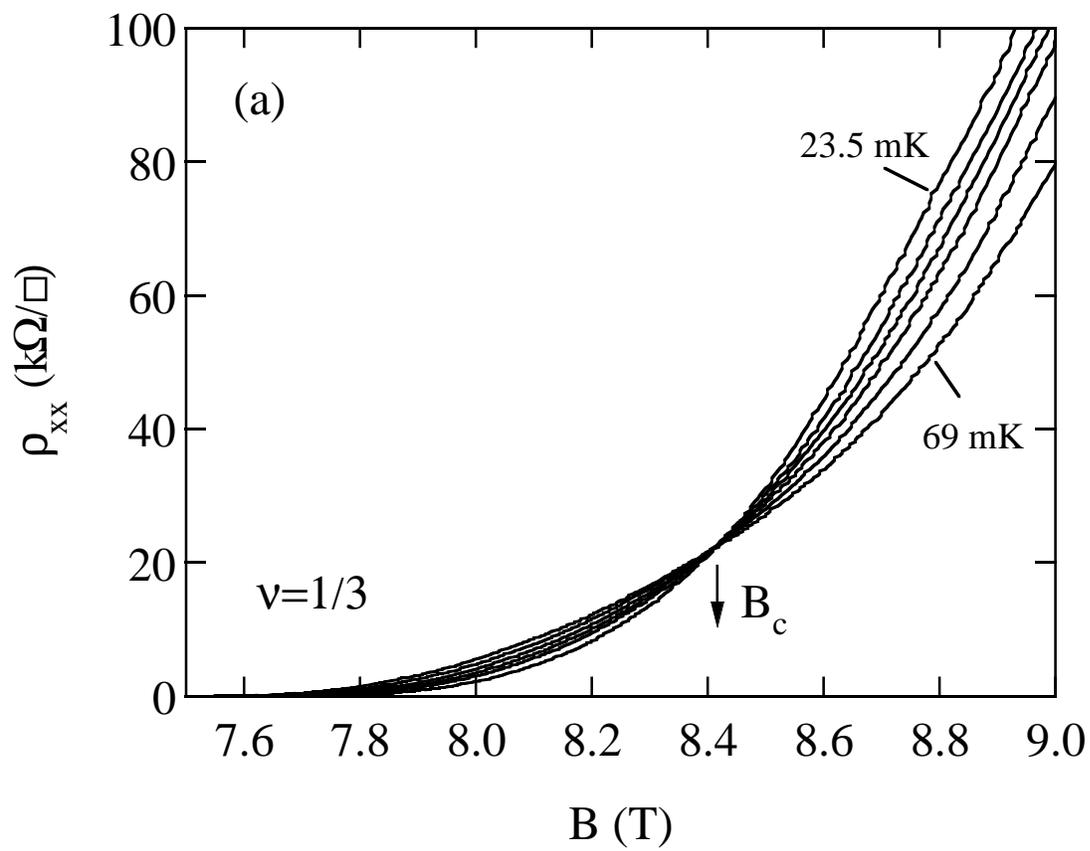
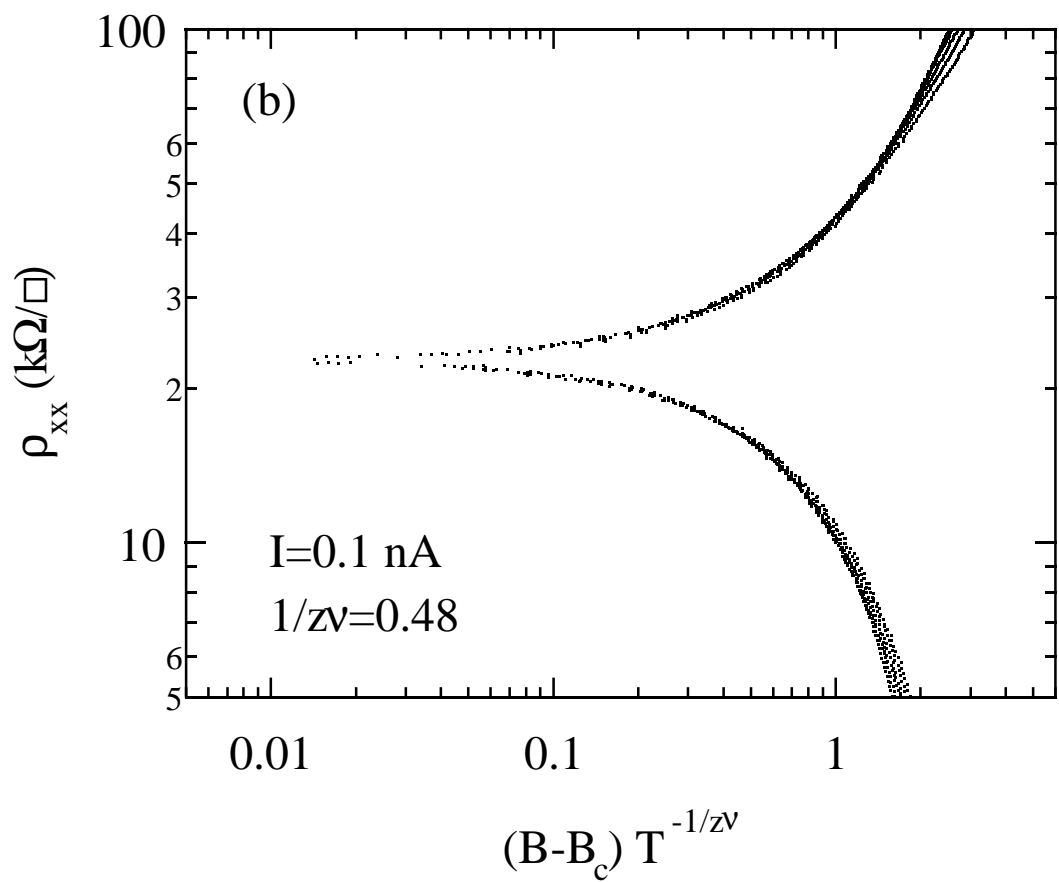
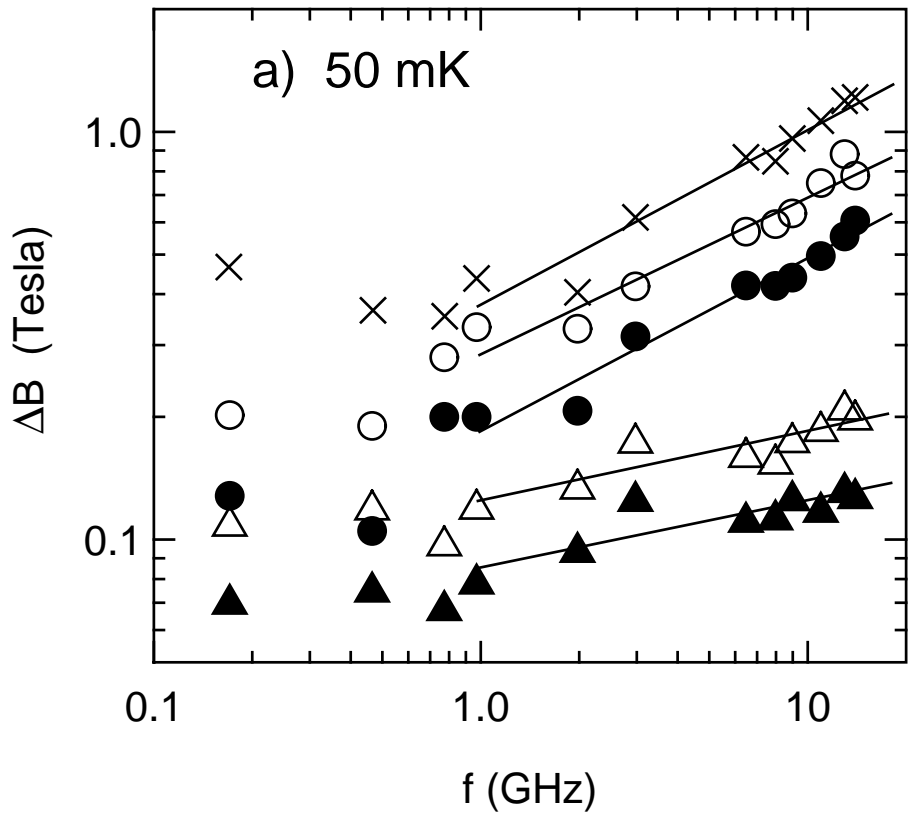
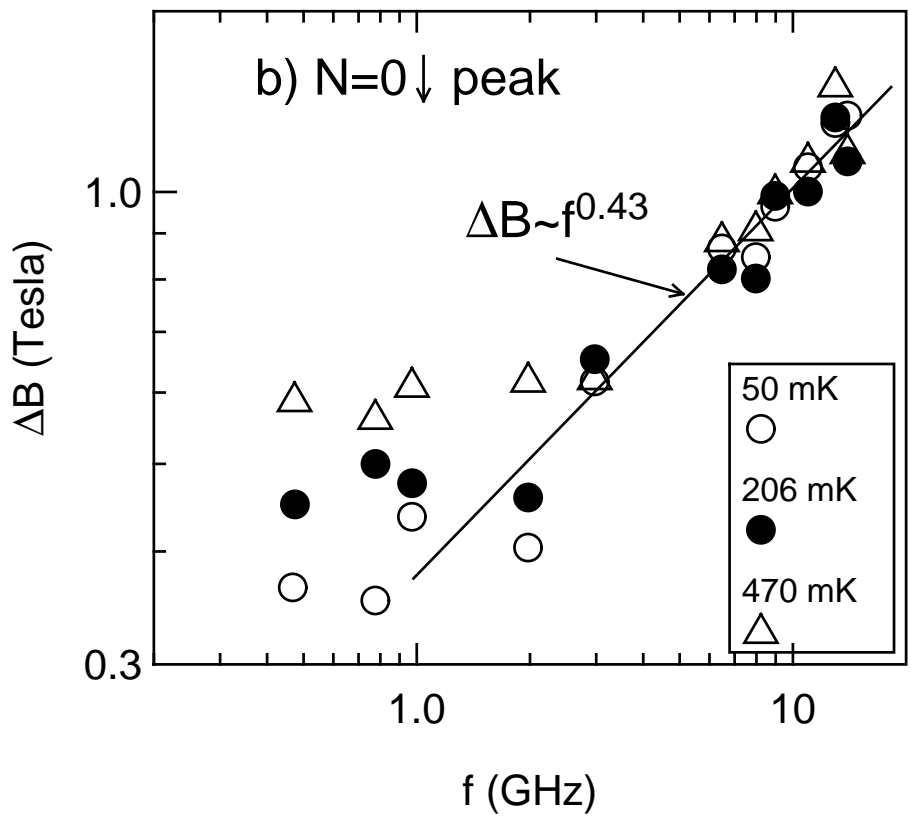
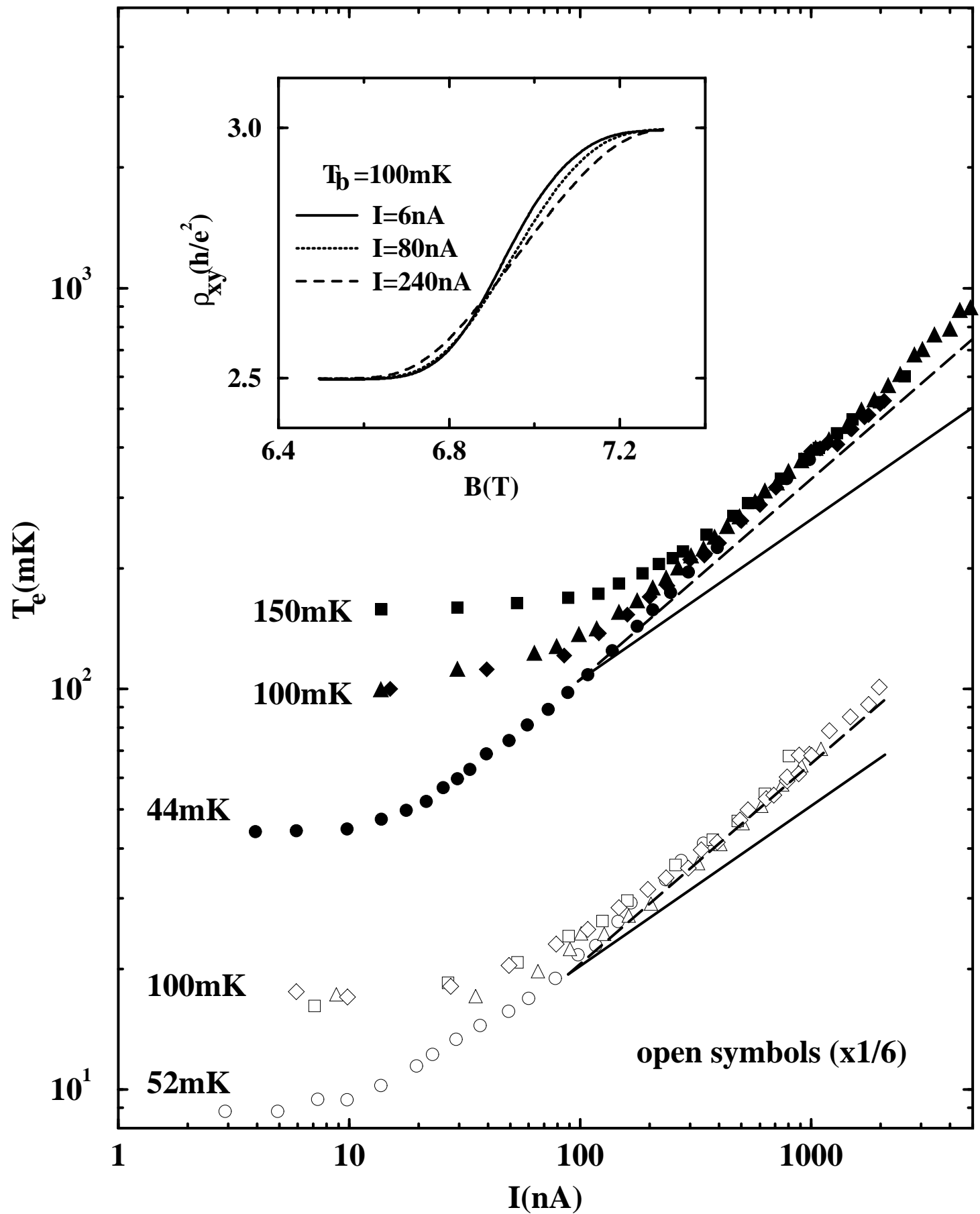
\end{document}